\documentclass[sigconf]{acmart} 

\AtBeginDocument{%
 }

\copyrightyear{2025}
\acmYear{2025}
\acmConference[CHI '25]{CHI Conference on Human Factors in Computing Systems}{April 26-May 1, 2025}{Yokohama, Japan}
\acmBooktitle{CHI Conference on Human Factors in Computing Systems (CHI '25), April 26-May 1, 2025, Yokohama, Japan}
\acmDOI{10.1145/3706598.3713359}
\acmISBN{979-8-4007-1394-1/25/04}

\begin{document}

\title[Leveraging AI-Generated Emotional Self-Voice to Nudge People towards their Ideal Selves]{Leveraging AI-Generated Emotional Self-Voice \protect\\ to Nudge People towards their Ideal Selves}

\renewcommand{\shorttitle}{Emotional Self-Voice}

\author{Cathy Mengying Fang}
\email{catfang@media.mit.edu}
\affiliation{%
 \institution{MIT Media Lab}
 \city{Cambridge}
 \country{USA}
}
\author{Phoebe Chua}
\email{phoebe@ahlab.org}
\affiliation{%
 \institution{Augmented Human Lab, National University of Singapore}
 \city{Singapore}
 \country{Singapore}
}
\author{Samantha Chan}
\email{samantha.chan@ntu.edu.sg}
\affiliation{%
 \institution{Nanyang Technological University}
 \city{Singapore}
 \country{Singapore}
}
\affiliation{%
 \institution{MIT Media Lab}
 \city{Cambridge}
 \country{USA}
}

\author{Joanne Leong}
\email{joaleong@media.mit.edu}
\affiliation{%
 \institution{MIT Media Lab}
 \city{Cambridge}
 \country{USA}
}
\author{Andria Bao}
\email{andria@mit.edu}
\affiliation{%
 \institution{MIT}
 \city{Cambridge}
 \country{USA}
}
\author{Pattie Maes}
\email{pattie@media.mit.edu}
\affiliation{%
 \institution{MIT Media Lab}
 \city{Cambridge}
 \country{USA}
}

\renewcommand{\shortauthors}{Fang et al.}

\begin{abstract}
Emotions, shaped by past experiences, significantly influence decision-making and goal pursuit. Traditional cognitive-behavioral techniques for personal development rely on mental imagery to envision ideal selves, but may be less effective for individuals who struggle with visualization. This paper introduces Emotional Self-Voice (ESV), a novel system combining emotionally expressive language models and voice cloning technologies to render customized responses in the user's own voice. We investigate the potential of ESV to nudge individuals towards their ideal selves in a study with 60 participants. Across all three conditions (ESV, text-only, and mental imagination), we observed an increase in resilience, confidence, motivation, and goal commitment, and the \textit{ESV} condition was perceived as uniquely engaging and personalized. We discuss the implications of designing generated self-voice systems as a personalized behavioral intervention for different scenarios.

\end{abstract}


\begin{teaserfigure}
\centering
 \includegraphics[width=.85\textwidth]{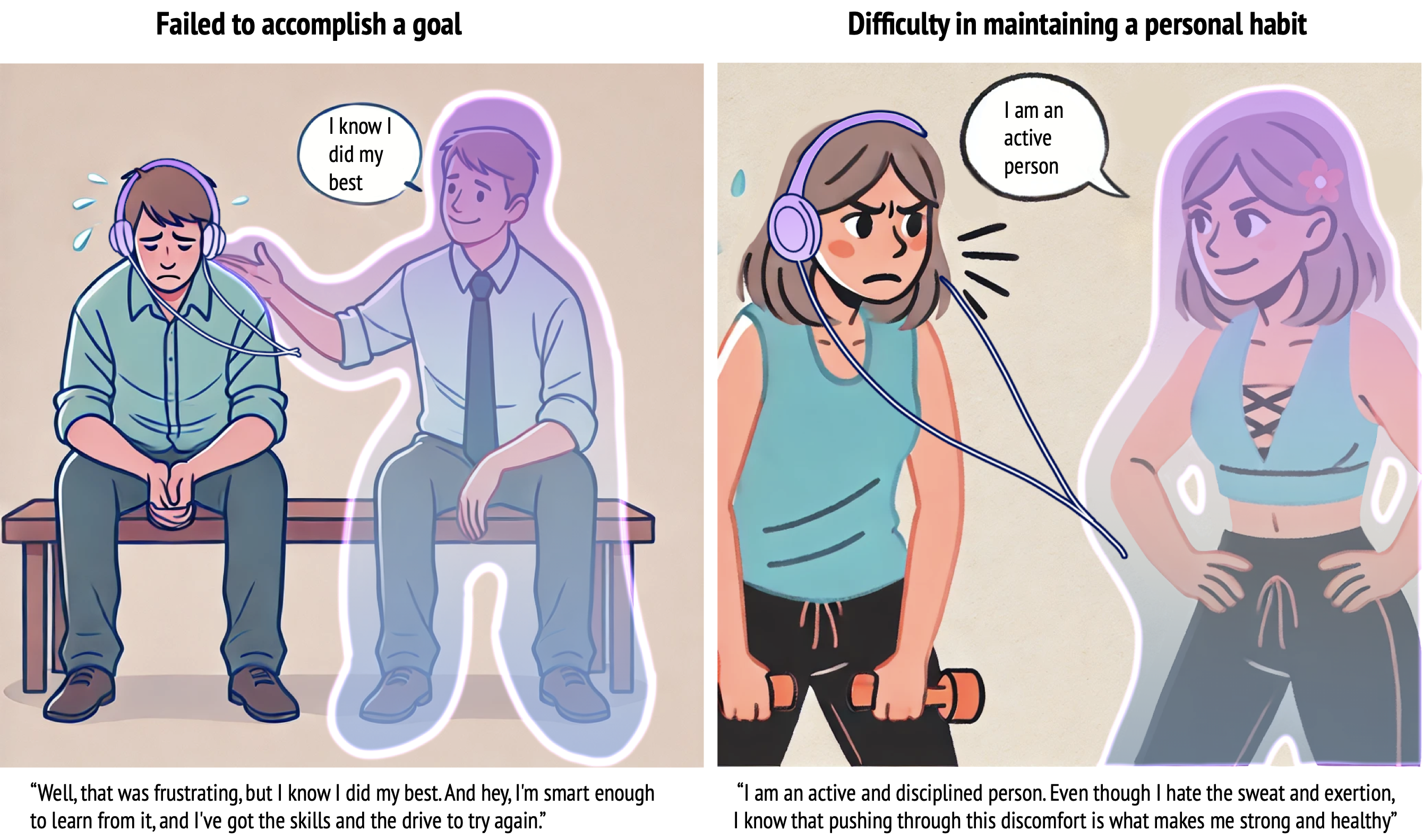}
 \caption{The Emotional Self Voice (ESV) system renders emotionally expressive ideal-self responses to difficult situations in the user's own voice using emotionally expressive language models and voice cloning technologies. Left: The individual hears his ideal self reframing failure as a learning opportunity. Right: Her ideal self reinforces a habit as part of her identity for motivation.}
 \Description{xxxx}
 \label{fig:teaser}
\end{teaserfigure}

\maketitle

\section{Introduction}

Emotion acts as a manifestation of our predictions based on prior experiences \cite{barrett2017theory}. As a result, emotion affects how we make decisions. For example, a positive outlook on the future can foster resilience and grit, enabling us to persevere in the face of challenges, but a pessimistic perspective may deter us from pursuing our goals. Relatedly, he concept of an ideal or aspirational version of one’s future self, rooted in self-discrepancy theory, has been shown to be a key driver of intrinsic motivation \cite{higgins1994ideal}. It provides a powerful framework for understanding goal-directed behavior to reduce the perceived gap between one’s current or actual self and the ideal self by making decisions consistent with one's goals \cite{higgins1994ideal}. In the same vein, research has shown that the vividness of the ``future self'' can promote future-oriented behaviors and reduce the gap between intention and behavior \cite{hershfield2011increasing, rosch2022promoting}. Traditional cognitive-behavioral therapy (CBT) and reframing practices aim at helping individuals set goals and receive positive reinforcement for their efforts often encourage individuals to create mental imagery of their alternative versions of themselves in a scenario. However, producing compelling mental images (including images, voice, and other sensory modalities) can be difficult, and some lack the ability altogether \cite{andrade2014assessing}. The question is then how to nudge an individual towards their ideal self. Concretely, the challenges lie in one's ability to experience an ideal version of oneself that is temporally and emotionally distant from their current self.

Recent advancements in generative models offer unique opportunities to address these challenges. Generative models can enable hyper-realistic visual-audio rendering of alternative versions of the self \cite{pataranutaporn2021ai, leong2021exploring,leong2023picture,leong2023using}. These experiences help individuals to take different perspectives or embody a different version of themselves. 

Voice presents a particularly intriguing avenue for exploration: common phrases like ''inner voice'' and ''voices in one's head'' reflect people's natural tendency to mentally simulate thoughts as speech, and hearing one's voice elicits higher arousal and alertness in a way that may be beneficial for goal achievement \cite{payne2021perceptual, kim2024myvoice}. The advent of large language model (LLM) enabled chatbots that can converse with users in natural language through both text and speech also opens up compelling possibilities for their positive application in behavioral interventions.


We introduce Emotional Self Voice (ESV), a novel system that combines emotionally expressive LLMs with voice cloning technologies to generate speech in various attitudes that sound like the user's own voice. ESV allows users to hear their ideal selves articulate thoughts and affirmations, rather than merely imagining them. Unlike previous studies that used static voice recordings, our approach leverages generative models to create dynamic, emotionally expressive speech that matches given attitudes while maintaining the user's voice characteristics.

Our research aims to investigate how hearing one's own voice respond to challenging situations with different attitudes affects self-perception and behavior for goal achievement. We hypothesize that: 
(1) Hearing an ideal-self response through ESV is more motivating and confidence-boosting than text-only or mental-imagination-only conditions, despite the potential initial discomfort with hearing one's own voice; 
(2) ESV enables people to more vividly imagine their ideal-self saying the response; 
(3) ESV leads to greater improvements in affect, resilience, and goal commitment compared to other conditions.

To investigate the potential of ESV, we conducted a between-subjects study with 60 participants with random assignment to one of three conditions: ESV, a text-only variant condition (seeing the generated ideal-self response in text without hearing the generated self-voice), or control (only mental imagination of the ideal-self response). The study focused on two scenarios related to goal achievement: overcoming failure in achieving a past goal and experiencing resistance in establishing a habit.

Our results reveal several key insights:

\begin{enumerate}
  \item Overall, the emotional valence ratings became significantly more positive (p<.001) across most conditions for both scenarios, with the \textit{ESV} condition having the most positive post-intervention valence ratings. Across all conditions, we observed an increase in resilience, confidence, motivation, and commitment, with the \textit{ESV} condition again ranked the highest for post-intervention confidence and motivation.
  \item Qualitative results show that \textit{ESV} condition has the most positive change in affective states (indicated by the sentiment analysis of post-intervention responses). The ideal-self concept, as presented through ESV, demonstrated the potential to enhance the vividness of future-oriented thinking and goal representation.
  \item The study revealed nuanced differences between responses towards past-focused unmet outcomes (scenario 1) and future-focused, progress-driven (scenario 2), highlighting the need for different temporal and motivational framing in behavioral interventions.
  \item Participants showed a preference for positivity and expressiveness in generated responses, balanced with realism, particularly in future-oriented scenarios.
\end{enumerate}

Our contributions are:
\begin{itemize}
  \item We developed a novel proof-of-concept system that generates Emotional Self-Voice---emotionally expressive responses spoken in the user’s own voice---to help people more vividly embody their ideal self in response to challenging situations.
   \item Our system combines generative text-to-speech and speech-to-speech models to successfully generate highly realistic, emotionally expressive self-voices, as validated by our technical characterization and participant feedback.
   \item Our user study delved into two scenarios of goal achievement and investigated how ESV affects people’s attitudes towards these personal experiences.
   \item Our approach is based on the self-discrepancy theory and we investigate how the concept of an ideal self can be used as an intervention for behavior change. We showed personalized emotional voices can serve as a novel mechanism for bridging the gap between actual and ideal self-perceptions.
   \item We also discuss design implications that can be generalized to future self-voice systems as a form of personalized behavioral intervention.

\end{itemize}

\section{Background \& Related Work}

\subsection{Towards the Ideal Self}
The ideal self is defined as ``a personal vision, or an image of what kind of person one wishes to be'' \cite{boyatzis2006ideal}. A breadth of positive psychology research suggests that the ideal-self concept serves as a key driver of intrinsic motivation to achieve personal and professional goals \cite{higgins2024placing, yeager2020can}, by providing a positive reference value that guides regulatory systems to eventually align one's current or actual self with the ideal, aspirational self \cite{higgins1994ideal}. Developing ideal-self concepts implicitly suggests that the current self is malleable and that qualities of the desired self can be developed. This perspective closely aligns with research on growth mindsets \cite{dweck2019mindsets}, which finds that the belief that one's abilities are not fixed but can be improved over time promotes resilience and achievement \cite{yeager2012mindsets}. Apart from positive references, self-regulation can also occur through avoidance of undesired end states: for example, commitment devices such as financial penalties and social accountability have shown a positive relationship with goal achievement \cite{lee2021sticky}. However, prior work has shown that the ideal self tends to show a stronger positive association with desired end states than a negative association with undesired end states, suggesting that movement toward the ideal self benefits from approach-oriented strategies than avoidance-based ones \cite{higgins1994ideal}.

Existing research on interventions for behavior change and goal achievement has primarily explored their potential to support goal capture, goal monitoring, and maintenance of goal-directed motivation \cite{lolla2023evaluating, duckworth2007grit}. Despite the popularity of such interventions, with over 1,300 apps related to goal achievement available on the Google Play Store, they often suffer from high attrition and limited user engagement \cite{lazar2015we}. To address these issues, \citet{dominick2020goals} proposed the incorporation of goals into one’s identity and found that framing a goal as part of one’s identity (e.g., ``I will think of myself as someone who eats healthy and works out.'') promotes goal-consistent choices to a greater extent than simply setting the goal (e.g., ``I will be mindful about what I eat.''). Reflection has also emerged as an important component of behavior change interventions \cite{cho2022reflection}. Applications for technology-mediated reflection, such as reflective conversational agents \cite{li2023exploring}, tools for recording and reflecting on daily activities \cite{isaacs2013echoes} and LLM-assisted journaling \cite{nepal2024contextual, song2024exploreself, kim2024mindfuldiary} not only facilitate goal monitoring, but also helps users see positive aspects of initially negative experiences, and draw lessons for future behavior.

In this work, we explore the use of reflection to prompt AI-generated media that nudges individuals towards their ideal selves by helping them to more vividly envision what the ideal self might sound like. In the next section, we discuss related works in the realm of AI-generated synthetic selves.

\subsection{AI-generated Synthetic Selves for Behavior Change}

Digital representations of oneself can alter our behavior. The Proteus effect \cite{yee2007proteus} is a phenomenon in which people's behaviors adapt to the traits of their avatars in virtual worlds. In virtual reality (VR), it has been shown that embodying different characters can influence people's cognition and behavior. For example, men who embodied Einstein in VR performed better on a cognitive task \cite{banakou2018virtually}, and embodying a young child-like avatar in VR can lead to changes in perception of object size and speaking behaviors \cite{tajadura2017embodiment}. In a study done by Osimo et al., subjects embodied Sigmund Freud and offered themselves counseling, capitalizing on the illusion of body ownership of the Freud body \cite{osimo2015conversations}.

Beyond static avatars, recent advancements in machine learning and generative AI technologies have made it possible to produce synthetic self-similar media for the purposes of wellbeing, learning, and skill development \cite{pataranutaporn2021ai, leong2023using, danry2022ai}. Researchers, for instance, have explored improving engagement in exercise \cite{clarke2023fakeforward} or developing public speaking skills by generating personalized videos of oneself speaking confidently in public speaking scenarios \cite{clarke2023fakeforward,leong2021investigating}. In the context of human-AI collaboration, studies find that virtual characters with self-similar appearance are perceived as more intelligent and elicited a stronger sense of co-presence, while those with self-similar voice are perceived as more likable and believable \cite{guo2024collaborating}. Synthesized future representations of oneself have also been shown to improve people's tendency to save for retirement \cite{hershfield2011increasing}, and chatting with an AI-generated future version of oneself can positively influence emotions and a feeling of self-continuity \cite{pataranutaporn2024future}. In contrast to generated videos, augmented reality (AR) filters have been shown as a means for real-time interventions. For example, AR filters making oneself look more inventor- or child-like has been shown to possibly boost creativity \cite{leong2021exploring}. In another example, it was found that the private application of filters on oneself in online public speaking situations can be a means to mitigate public speaking anxiety and encourage further development of the skill \cite{leong2023picture}. 


More related to our work are the explorations into synthetic voices. \citet{costa2018regulating} found that playing back an altered version of one's voice to be calmer made people feel less anxious during relationship conflicts, whereas lowering the voice made people feel more powerful during debates. Explorations have also been made into using the voice of oneself \cite{kim2024myvoice} or familiar people (e.g., family and friends) to improve the efficacy of notifications \cite{chan2021kinvoices}. Next, we discuss prior work that use voice as a behavioral intervention and the generation of emotionally expressive voices.

\subsection{Voice-based Intervention for Emotion Regulation}

Voices not only express feelings but can also affect one's feelings \cite{aucouturier2016covert,hatfield1995impact}. Hearing the playback of one's own voice is not common. Most people find it unpleasant \cite{holmes2018familiar, kim2012visualizing} and it could cause attentional bias~\cite{daryadar2015effect}. People also had negative affective and defensive reactions to hearing one's voice due to the mismatch in what they expected to hear and what was heard~\cite{holzman1966voice,holzman1966listening}. Using one's own voice as an intervention has a few favorable aspects. The presence of self-relevant information in a voice, such as self-similarity and self-generation, has been linked to enhanced speed and accuracy of voice perception \cite{rosi2024effects}. Perceptually, hearing one's voice elicits higher arousal and alertness \cite{payne2021perceptual}, and are possibly more familiar than other voices~\cite{xu2013acoustic}. Self-voices also seem to be processed implicitly, even when they are unrecognized by their owners~\cite{daryadar2015effect}. Psychologically, there are a few effects that work in the favor of using self-voice. People tend to process self-related information more importantly and positively due to the self-referencing effect (e.g., when the information states things about or related to ``you'')~\cite{rogers1977self}. Self-referencing messages have been shown to improve involvement and intention to increase physical activity and healthy eating~\cite{wirtz2020does}. Many voices seem familiar due to frequent exposure or the ``mere exposure'' effect that influences personal preferences and enhances affect-based trust~\cite{kwan2015mere}. Familiarity of voices induces more attention, engagement, and recall \cite{holmes2018familiar,chan2021kinvoices, england2021effects}. People might prefer to interact with others (human and digital systems) who are similar to themselves due to the similar-attraction effect \cite{reeves1996media,nass1995can}. These might help strengthen one's acceptance and embodiment of the emotional self-voice of the ideal self.

In HCI and education research, people reacted most quickly to self-voice notifications compared to familiar voices (of instructors and lab mates) and unfamiliar voices~\cite{bhatia2006listening}. Video game avatars with self-similar voices elicited greater identification, and improved performance and immersion in the game \cite{kao2021effects}. Self-voice alarms from smartphones produced discomfort but better vocabulary learning task completion and goal achievement compared to other voice alarms~\cite{kim2024myvoice}.
Echoing one's self-voice during learning improves recall~\cite{lane1986learning} and developing a second language self-voice improves language learning~\cite{tomlinson2001inner}.

However, most of the prior work used individuals' recorded voices. Only a few studies have explored the digital altering of self-voices to influence emotions. A key example was a study that generated a calmer version of one's voice, which lowered anxiety~\cite{costa2018regulating}. The effects of synthesized versus recorded voices' effect on emotion and perceptions of content have been studied~\cite{nass2001effects}, showing that happy synthesized and recorded voices made content seem happier while sad ones made content seem less happy. People also liked the content more and found the information more credible when the emotion of the voice and the content were matched. 

More recently, the use of generative AI models for creating emotionally expressive voices has surged, and the models have become increasingly more capable of generating voices with a wide emotional range. Models like EmoCtrl-TTS \cite{wu2024laugh} and VAW-GAN \cite{zhou2021seen} can produce time-varying emotional states within the generated speech~\cite{triantafyllopoulos2023overview}. There is a growing number of open-sourced \cite{Eren_Coqui_TTS_2021,jia2018transfer} and proprietary models\footnote{\url{https://themetavoice.xyz/}} \footnote{\url{https://speech.fish.audio/}} that enable zero-shot voice cloning.


Researchers have not thoroughly studied the effects of AI generated \textbf{self voices} on emotions and behaviors. Leveraging these emerging models' capabilities, we can allow the speech (both the content and the voice) to be dynamically generated instead of pre-recorded voices, and we can generate synthetic voices with the emotional expressiveness of natural voices. This is an extension of ``voice as a design material''~\cite{sutton2019voice} where we explore not only the voice characteristics but the combined effect of emotional voice and self voice as an intervention.

In our work, we leverage the latest models for synthesizing both the content from the perspective of an ideal self as well as the spoken voices in the style of the user's voice. To the best of our knowledge, this work is the first to ever explore the development and effect of using synthetic emotional, self-voice on individuals' perception and behavior.


\section{Generated Emotional Self Voice for Nudging towards Ideal Self }
In this work, we use the phrase ``self-voice'' to refer to authentic and generated versions of one's own voice, and ``Emotional Self Voice'' (ESV) to refer to our proposed system and its generated outputs. We designed Emotional Self Voice to allow individuals to hear an ideal version of themselves responding to a situation in which they had either failed to achieve a personal goal, or were struggling to persist in achieving some goal. These responses aim to nudge individuals toward their ideal selves by presenting a vivid mental picture of what that self might sound like. In the following subsections, we describe the implementation of the Emotional Self Voice system in further detail.

\subsection{System Implementation}
\begin{figure}[t!]
  \centering
  \includegraphics[width=1\linewidth]{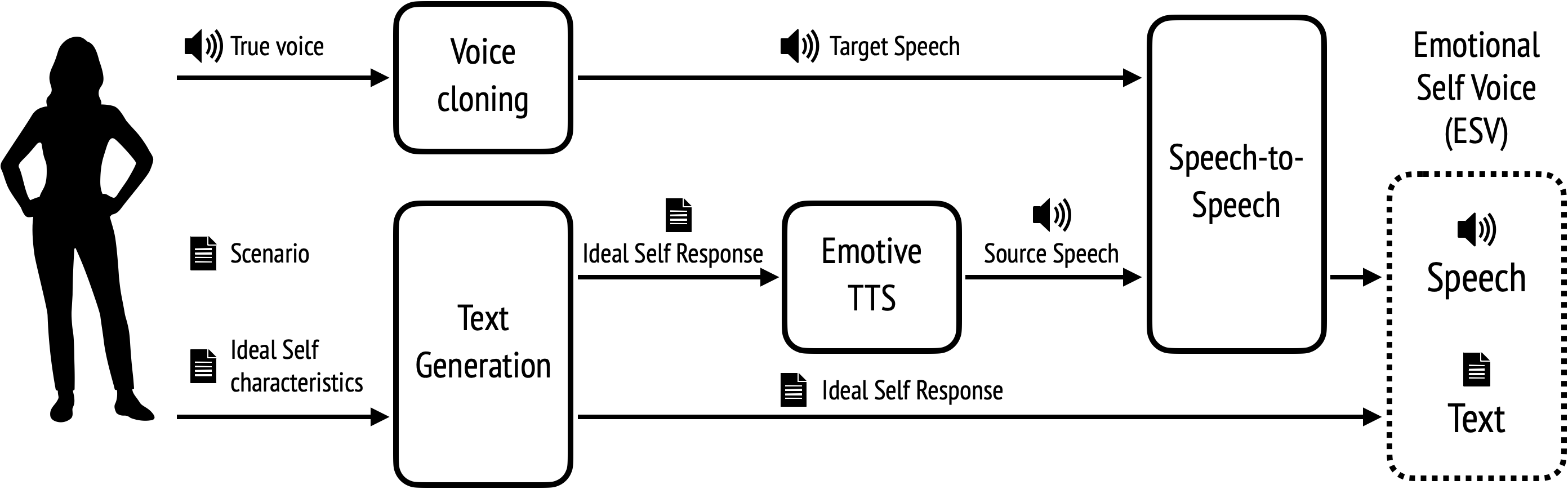}
  \caption{An overview of the Emotional Self Voice system. The system consists of multiple generative models for synthesizing texts and speeches. The individual provides a scenario and the corresponding ideal self characteristics. Our system generates the text and audio response of in the style of the ideal self and in the voice of the individual.}
  \label{fig:system}
  \Description{xxx}
  \vspace{-8px}
\end{figure}

Figure \ref{fig:system} overviews our system, which has three main components: the generation of a hypothetical response of the ideal self, the generation of an emotionally expressive speech based on the content, and finally the cloning of the generated speech in the style of the individual's own voice to achieve the self-voice effect.

\subsubsection{Step 1: Text Generation of Ideal Self Responses} 
\label{sec:text-prompt}
The goal of the text generation step is two-fold: (1) to represent the personality and values of one's ideal self, and (2) to generate cues for emotionally expressive speech. We first asked participants to describe scenarios in which they had failed to achieve a personal goal, or were struggling to establish a new personal habit. Then, we asked them about the characteristics of their ideal self in the provided scenario. We used GPT-4o, an LLM, to generate ideal-self responses based on a scenario provided by the user. In describing the scenario and characteristics of their ideal self, users essentially completed an exercise akin to a journaling prompt for self-reflection. However, the LLM-generated text exposes the user to possible responses that they might have not thought of. Finally, we also prompted the LLM to add vocal bursts (such as sigh), discourse markers (such as ''well'', ''I mean'', ''uh''), and other natural emotive embellishments to the generated responses to facilitate the generation of natural and emotionally expressive speech in the next step of our system. The full prompt can be found in the Appendix \ref{sec:appendix-text-prompt}.

\subsubsection{Step 2: Emotional Speech Generation from Text}

The emotional content of a speech comes not only from its linguistic content but also \textit{how} it is spoken --- from pitch and intonations to non-verbal expressions \cite{cutler1997prosody}. Hence, in this step, we used Hume AI’s empathic voice interface\footnote{\url{https://dev.hume.ai/docs/empathic-voice-interface-evi/overview}} to generate emotionally expressive speech based on the output of Step 1. Although Hume AI and new text-to-speech models (TTS) \cite{wu2024laugh} represent an advancement over previous TTS models that were only able to generate monotonic or emotionally neutral outputs\footnote{\url{https://openai.com/index/chatgpt-can-now-see-hear-and-speak/}}, at the time of writing, the model’s API does not allow one to directly adjust the emotional parameters of the speech output. Thus, the emotional quality of the generated speech largely depends on the expressiveness of the text. In a later section, we present the evaluation of the emotional quality of the generated speech, which shows the reliability and expressiveness of the generation.

\subsubsection{Step 3: Voice Cloning and Generating Emotional Self-Voice with Speech-to-Speech models}
Increasingly, there are many open-source models \cite{kang2022end,betker2023better,wang2023neural} and commercial models \cite{coqui,elevenlabs} that allow zero-shot and speech-to-speech generation with a target voice. In choosing the model to perform the voice cloning and speech-to-speech generation, we prioritize stability and speed of the generation, since the focus of this work is on the effect of an ESV system on individuals' perception and behavior. Most open source models are unfortunately both computationally expensive and slow while the output quality is not convincing. In the end, we chose ElevenLab's voice cloning API\footnote{\url{https://elevenlabs.io/docs/voices/voice-lab/instant-voice-cloning}} given its speed and quality. 

We experimented with different text for the users to read for voice cloning, and we referenced best practices to achieve the best generation quality\footnote{\url{https://elevenlabs.io/docs/voices/voice-lab/instant-voice-cloning}}. Our initial tests with neutral text produced a more monotone voice, therefore, we settled on a set of emotionally positive and negative texts to elicit a broad range of emotional expressions (see Appendix \ref{sec:appendix-cloning}).

Once a reference of the user's voice is obtained, the next step is to create a transformation between the source speech (i.e., generated emotional speech from the previous component) and the target voice (i.e., the user's voice). Initially, we also experimented with ElevenLab's text-to-speech API\footnote{\url{https://elevenlabs.io/docs/api-reference/text-to-speech}} and used the generated text from Step 1. The result yielded monotonic speech and the quality varied depending on the emotional tone of the target user voice. Thus, we use the emotionally generated speech from Step 2 as the source speech input for the speech-to-speech API. We also tuned different parameters of the API such that the output preserves both the emotional range of the source speech and the characteristics of the target voice.

\begin{figure}[t!]
  \centering
  \includegraphics[width=0.75\linewidth]{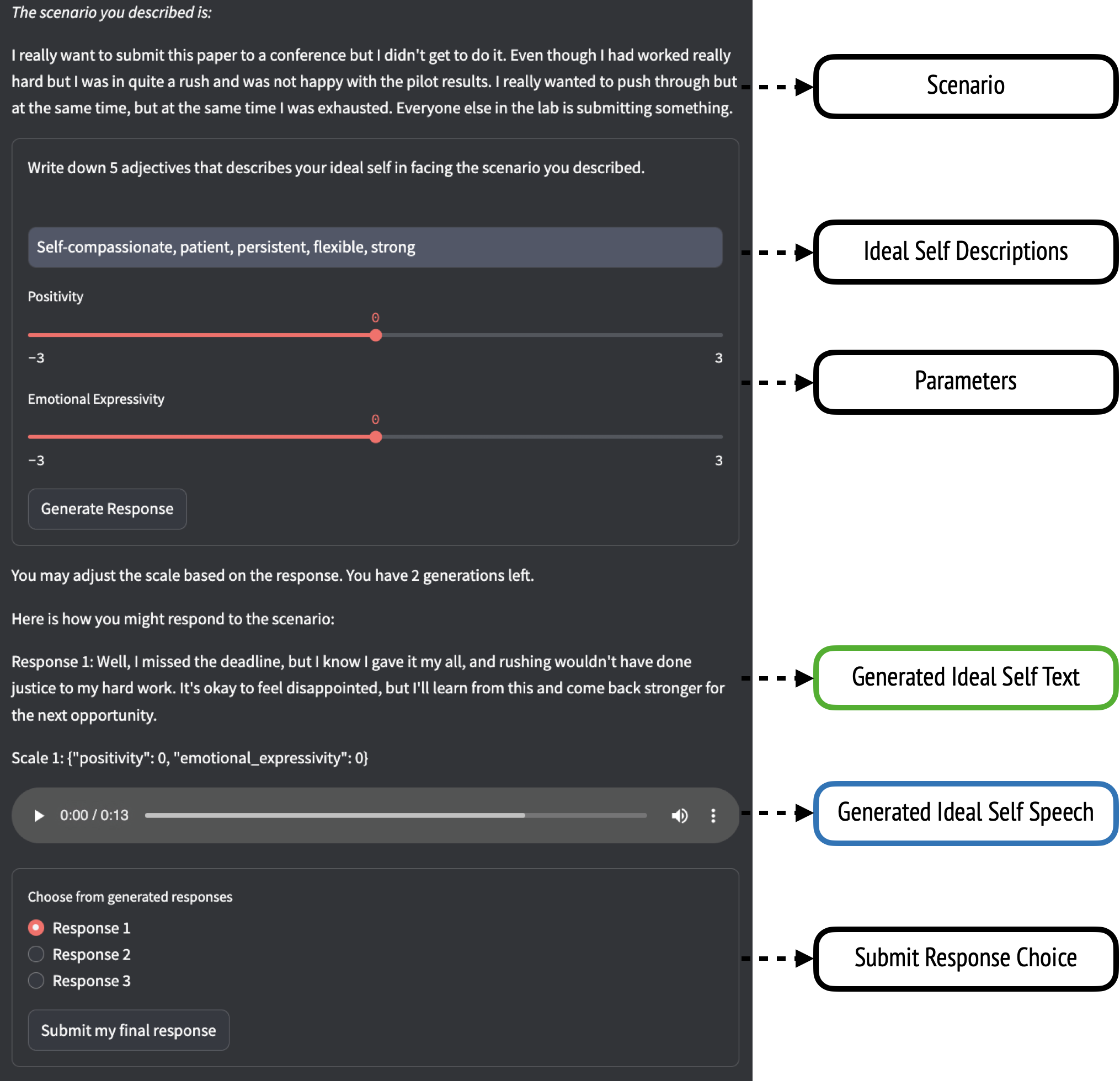}
  \caption{The user interface of the Emotional Self Voice system for the study. The top shows the scenario provided by the user. The user is asked to provide adjectives that describe the ideal self and then receives the generated text and self-voice response. The user can fine-tune the response by adjusting the parameters.}
  \label{fig:interface}
  \vspace{-10px}
\end{figure}

\begin{figure*}

  \centering
  \includegraphics[width=1\textwidth]{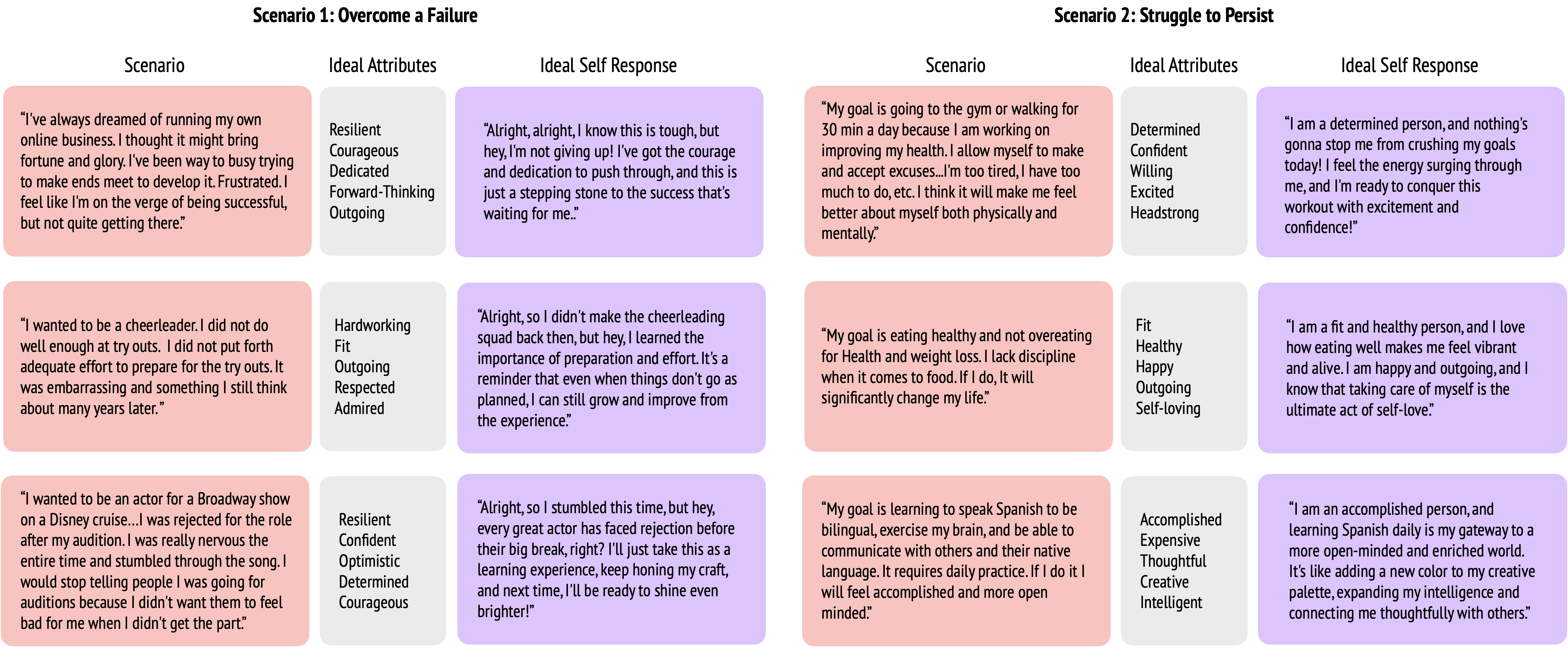}
  \caption{Example scenarios, ideal self attributes and generated ideal self responses for both scenarios.}
  \label{fig:example}
  \vspace{-10px}
\end{figure*}

\subsubsection{Final System Setup}
Putting everything together, the final system implementation is as follows: we used ElevenLabs' instant voice cloning API\footnote{\url{https://elevenlabs.io/docs/voices/voice-lab/instant-voice-cloning}} to synthesize a clone of self-voice clips based on short voice samples of about 1 minute, which returns a unique voice ID. We then used GPT-4o\footnote{\url{https://platform.openai.com/docs/models/gpt-4o}} (temperature=0.15) to generate a text response based on a given scenario and set of personalities. In order to generate emotionally expressive speech output, we use Hume AI's LLM (EVI v1\footnote{\url{https://dev.hume.ai/docs/empathic-voice-interface-evi/overview}}) to perform text-to-speech. We set the gender of the voice configuration to match the gender of the user. The generated voice clip is then inputted as the source speech to ElevenLab's speech-to-speech model\footnote{\url{https://elevenlabs.io/docs/speech-synthesis/models}} (Multilingual v2, stability=1.0, similarity\_boost=0.7, style=0) to translate the generated speech into the user's voice (target voice). The end-to-end generation time is about 30 seconds/sentence from generating the input text to generating the emotional self voice.



\section{User Study}
Our research interest is whether ESV can be used as an intervention to nudge individuals toward their ideal self. We start by describing the setup of the study such as the scenarios that we used as case studies and elaborate on the relevant technical adaptations for the purpose of the study, and then we describe the procedure and measurements.

\subsection{Scenarios} \label{sec:scenario}
We chose scenarios within self-development, where individuals set out to achieve some goal. We did not consider situations where ESV could help interpersonal relationships. Although important, for this study, we chose not to focus on situations where other individuals' perception of an event has a significant effect, e.g., asking the boss for a raise, where the relationship between the boss and the individual is unknown. We focused on situations where the individual's perception of the event has a significant effect on consequential decisions and behaviors (e.g., having self-deprecating thoughts after a failure, or a lack of courage to pursue a goal due to imposter syndrome). In particular, we looked at two types of scenarios within the category of individual goal-achievement: (1) overcoming a past failure of accomplishing a goal, and (2) struggling to persist towards establishing a personal habit, which are common challenges that occur in the process of achieving a goal. 

While both scenarios involve challenges in behavior change, they differ fundamentally in terms of the following key aspects:

Temporal and cognitive focus---``Facing failure in achieving a goal'' is inherently retrospective. It involves reflection on past actions and outcomes, often accompanied by emotions like regret or disappointment. The cognitive process focuses on resilience, evaluating what went wrong, and learning from past experiences. ``Experiencing difficulty in establishing a habit'' is forward-looking. It emphasizes ongoing effort and the anticipation of future behavior, with a cognitive focus on persistence and motivation to form routines \cite{wood2007new}.

Motivation---Goal failure triggers a corrective motivation, prompting individuals to re-evaluate strategies or even adjust their goals \cite{yeager2020can}. Habit formation struggles elicit a maintenance motivation, focusing on repetition and consistency. 

Outcome vs. process orientation---Goals are typically tied to specific outcomes, while habits are associated with the process itself and focus on automaticity and repetition. A failed goal implies an unmet outcome, whereas a struggling habit suggests an ongoing process that may still succeed with continued effort \cite{clear2018atomic}.

For each type of scenario, we asked participants to provide a personal scenario based on our prompts so that the scenarios elicit real emotional responses. We also asked participants to describe their ideal self characteristics in the text. The scenarios and the characteristics were then provided to the LLM to generate a hypothetical ideal-self response to the scenarios. The motivation behind generating the text prompt is that people often might lack words to describe how they would ideally respond. Similar to how people confide in therapists or friends to hear how others might respond to their situation, providing generated responses might spark some inspiration. Figure \ref{fig:example} shows examples of scenarios, ideal self attributes provided by study participants and the corresponding ideal self response generated by our system.

We had an iterative approach to the prompts for generating the ideal-self response. There are two aspects we focused on. First, the perspective of the response is inspired by the cognitive-behavioral therapy practice, which we detail in the following sections. Second is the emotional expressiveness of the response. The generation of the audio output is based on the text prompt, of which the sentiment reflects the sentiment of the text response. For example, adding discourse markers makes the responses sound more natural and engaging as described in section \ref{sec:text-prompt}.

Next, we describe the prompts we used for the human participants and the LLM for generating the text.

\begin{figure*}

  \centering
  \includegraphics[width=.8\textwidth]{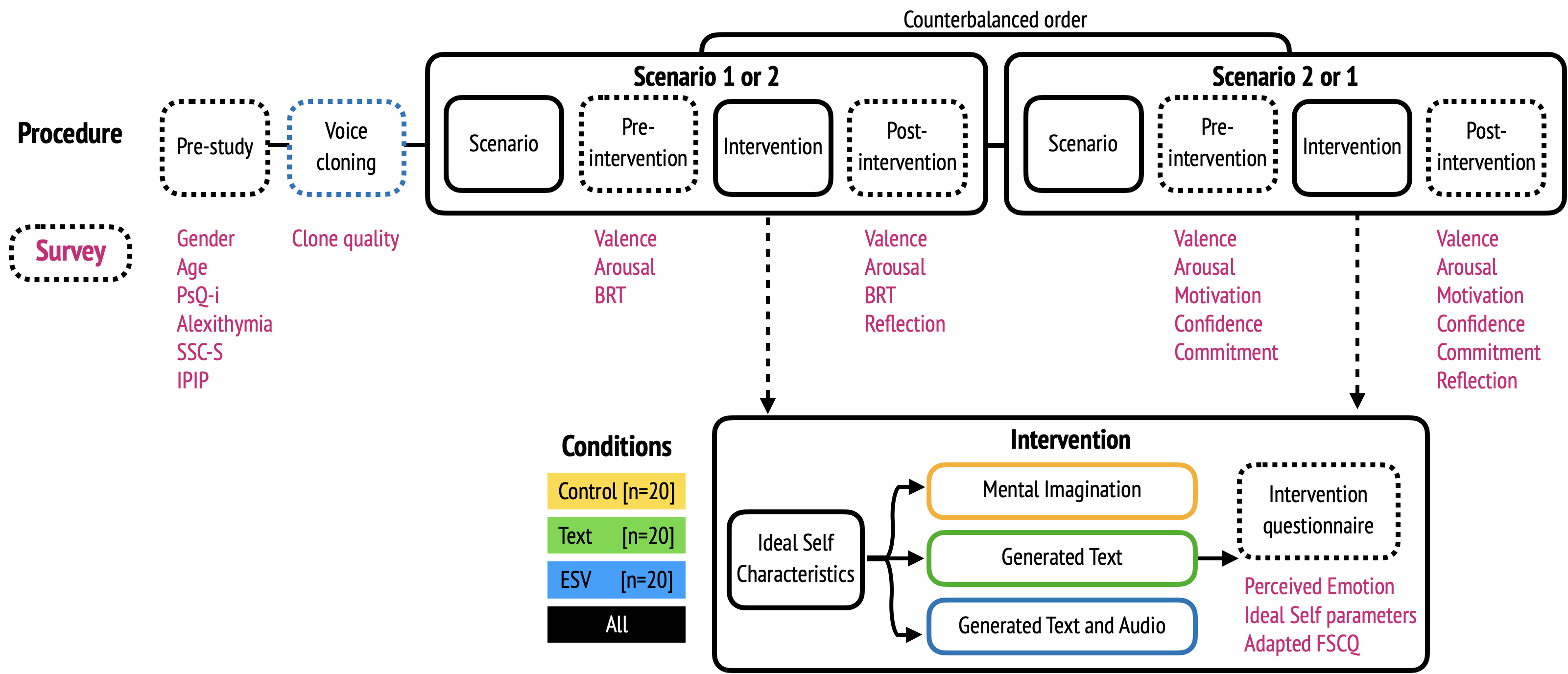}
  \caption{An overview of the user study procedure. PsQ-i: Plymouth Sensory Imagery Questionnaire; SSC-S: Self-Compassion Scale; IPIP: assesses the Big Five personality dimensions; BRT: Benchmark Resilience Tool; FSCQ: Future-Self Continuity Questionnaire.}
  \label{fig:procedure}
  \Description{xxx}
  \vspace{-10px}
\end{figure*}

\subsubsection{Scenario 1: Overcoming the Failure of Achieving a Goal}
In this scenario, we prompted participants to \textit{reflect and write about a time when they had a goal or something they wanted to accomplish but they didn’t succeed}. In particular, we asked: ``What was the goal?'', ``Why did you want to achieve it?'', ``What was the reason you didn't succeed?'', and ``How did that make you feel?''
In this situation of failure, one would reasonably feel discouraged, and worse, people with low self-compassion may attribute failure as a characteristic of themselves. The latter could exaggerate the severity of the event and deter future attempts at the goal. We adapted the common cognitive-behavior therapy strategy of reframing by replacing possible unhelpful thoughts with ones that bring out positive outlooks and constructive next steps \cite{wenzel2017basic}. The prompt we used for the LLM to generate the ideal-self text response is: ``Your task is to imagine yourself as the person with these trait personalities would say to themselves in the given scenario that would help them reframe the situation and be resilient in the face of failure.''

\subsubsection{Scenario 2: Struggling to Persist in Pursuing a Goal}
We prompted participants to \textit{describe a personal habit that they are having a difficult time establishing}. In particular, we asked: ``What is the habit?'', ``Why do you want to cultivate it?'', ``What is the reason it is challenging?'', and ``How do you think successfully forming the habit will make you feel?'' In this scenario, the individual needs motivation, confidence, and commitment to persist. 
Research has shown that individuals who think of goals as identities felt the goals were easier to pursue and pursued goal-consistent choices \cite{dominick2020goals}. We adapted the concept of ``goals as identities'' where the ideal self embodies someone with the goal already integrated as their identity. The prompt we used for the LLM to generate the ideal-self text response is: ``Your task is to imagine yourself as the person with these trait personalities would say to themselves in the given scenario that would encourage them to keep up with the habit. You should try to embody the person when their habit has become their identity. Use the template of ``I am a xxx person. ''

For the full prompt to the LLM for both scenarios, please see Appendix \ref{sec:appendix-text-prompt}.

\subsection{Conditions}
We are interested in how different aspects of the ESV system contribute to one's perception of an event. We employed three conditions: \textit{Control}, \textit{Text}, and \textit{ESV}.
The \textit{Control} condition is meant to mirror the internal imagination process one might have on their own or prompted by someone like a therapist or a friend. In this condition, the participants do not receive AI-generated text responses based on their ideal self and scenario, and they do not hear the generated self-voice saying the responses.
The \textit{Text} condition isolates the effect of self-voice. In this condition, the participants receive generated text responses but do not hear the generated self-voice saying the responses to them.
Finally, the \textit{ESV} condition goes through the voice cloning procedure and receives the generated response in both text and self-voice.
The study was designed to be between-condition to avoid the spill-over effect of experiencing multiple modalities of the intervention.

\subsection{Participants}
We recruited 64 participants from Prolific, and 4 were dropped from the data pool due to failing attention checks and giving incomplete responses. This left us with a final set of 60 participants equally split across three conditions. Note that while our sample size aligns with prior work investigating voice-based interventions \cite{chan2021kinvoices, kim2024myvoice}, and we acknowledge the exploratory nature of this study. Given the novelty of our specific research questions, there were no directly comparable studies from which to derive expected effect sizes. This work therefore serves as an initial investigation to estimate effect sizes and inform sample size planning for future confirmatory studies. The participants group consisted of 28 male (M), 30 female (F), and 2 non-binary (NB) individuals (\textit{Control}: 10M, 9F, 1NB; \textit{Text}: 8M, 11F; 1NB; \textit{ESV}: 10M, 10F). Participants ranged from age 18 and 56 (\textit{Control}: mean= 33.8, SD=7.53; \textit{Text}: mean=38.35, SD=8.60; \textit{ESV}: mean=35, SD=10.27). We provide a full breakdown of participants' demographics and other control variables per condition in Appendix Section \ref{tab:demographics}. All participants resided in the United States. They were screened for being at least 18 years old and speaking fluent English. Participants in the \textit{ESV} condition were screened for having normal or corrected to normal hearing. The participants spent an average of 32.4 minutes on the study, and the participants received \$15 for completing the study. The study was approved by our Institutional Review Board (protocol \#2408001387).

\subsection{Procedure}
Figure \ref{fig:procedure} overviews the study procedure. 
\subsubsection{Onboarding}
Participants provided informed consent to voice cloning and to have their survey data used anonymously prior to proceeding to the rest of the survey. Once enrolled in the study, participants provided their gender and age, and completed a questionnaire measuring Alexithymia which assesses difficulty in perceiving emotions (PAQ-S) \cite{preece2023perth}, the Plymouth Sensory Imagery Questionnaire which assesses the vividness of mental image (Psi-Q sound and feeling subscales \cite{andrade2014assessing}), and the Self-Compassion Scale (SSCS-S \cite{neff2021development}). They also completed the Mini-IPIP \cite{donnellan2006mini} (assesses the Big Five personality dimensions) twice, with respect to who they actually are and who they would ideally like to be. We then used the difference of the IPIP scores of the ideal and current self to calculate the amount of self-discrepancy \cite{stanley2015distance}.

\subsubsection{Voice Cloning}
Next, participants in the \textit{ESV} condition were asked to record an emotionally expressive sample of their voice by reading sentences aloud (Appendix \ref{sec:appendix-cloning}). They then listened to their true voice recording and their cloned self-voice reading an emotionally neutral text. They were asked about their perception of hearing their own voice and then rated the quality of the cloned voice in terms of familiarity, similarity, naturalness, and quality. The full questionnaire is in the Appendix \ref{sec:appendix-survey}.

\subsubsection{Intervention}
Participants were presented with a reflection task related to the two scenarios in a randomized order, counterbalanced within each condition. For each scenario, they were first asked to describe a scenario by answering questions described in Section \ref{sec:scenario}. 
During the intervention, all participants were asked to ``write down 5 adjectives that describe your ideal self in facing the scenario you described.'' After inputting the ideal-self attributes, the \textit{control} condition group was simply asked to ``imagine this ideal version of yourself with these personalities and what would your ideal self say to yourself in this scenario.'' Both \textit{Text} and \textit{ESV} conditions received generated text by the LLM, and the \textit{ESV} condition group also received the cloned self-voice saying the generated text. In addition, participants in these two conditions also were presented with two sliders that allowed them to adjust the generated responses a maximum of three times. The two sliders were \textit{positivity} and \textit{emotional expressiveness} on a scale of -3 to +3 (Figure \ref{fig:interface}). These two scales are common features of sentiment. The intention behind the sliders is two-fold: first, the participants might not be immediately satisfied with generated responses compared to how they envisioned what their ideal self might say. We wanted to provide an opportunity for fine-tuning the response. Second, we are also interested in sampling people's preferences to develop an emotional characterization of an ideal self. We were inspired by the Gibbs Sampling with People (GSP) approach \cite{harrison2020gibbs}, where in each iteration the participant uses a slider to manipulate a stimulus dimension to optimize a given criterion. At the end of the intervention, all participants were asked questions about the vividness, similarity, and preference of their ideal self. We adapted the self-continuity questionnaire (FSCQ \cite{sokol2020development}), where we replaced ``future self'' with ``ideal self''. We believe this adaption is valid as the ideal self is a particular version of the future self. They were also asked to describe the perceived emotions from the generated text and the generated audio (\textit{ESV} condition only).

Before and after the intervention, we asked the participants to fill out some questions about their affective state and aspects related to their goal. These questions are about our main outcomes, which we elaborate further in the following section.


\subsection{Research Questions and Measurements}
Given the two scenarios are emotionally negative aspects of goal-achieving, our main research questions are around the effect of ESV on individual's affect and psychological aspects around goals.

\textbf{H1: In a scenario of failing to achieve a personal goal, the Emotional Self Voice (ESV) intervention (a) induces more positive affect, and (b) more than the text-only condition and the control condition.}
To test this, we measured the difference in one's valence (1: unpleasant, 7: pleasant) and arousal (1: low intensity, 7: high intensity) using a 7-point scale, before and after the intervention. We also compared the difference in their explanation to their affective states pre- and post-intervention. 

\textbf{H2: In a scenario of failing to achieve a personal goal, ESV (a) increases an individual's resilience, and (b) more than the text-only condition and the control condition.}
To test this, we measured the difference in one's resilience to failure using the Benchmark Resilience Tool (BRT) scale \cite{smith2008brief} before and after the intervention. We also asked post-intervention what they then thought was the reason behind the failure and their action and how they felt about the failure.

\textbf{H3: In a scenario of having difficulty in establishing a personal habit, ESV (a) induces more positive affect, and (b) more than the text-only condition and the control condition.}
To test this, likewise, we measured the difference in one's valence and arousal, before and after the intervention. We also compared the difference in their explanation to their affective states pre- and post-intervention.

\textbf{H4: In a scenario of having difficulty in establishing a personal habit, ESV (a) increases an individual's confidence and motivation, and (b) more than the text-only condition and the control condition.}
To test this, we measured the difference in one's confidence and motivation with a questionnaire before and after the intervention. We also asked the participants post-intervention the reason for the difficulty behind establishing their personal habits and how they then felt about their ability to establish the personal habits.

\textbf{H5: In a scenario of having difficulty in establishing a personal habit, ESV (a) increases the individual's commitment to the goal, and (b) more than the text-only condition and the control condition.}
To test this, we measured the difference in one's goal commitment using the Goal Commitment Measure scale \cite{hollenbeck1989investigation} before and after the intervention. Again, we looked at the free text responses to the above questions post-intervention.

Note that, due to the short-term nature of the study, we were not able to measure and therefore are not claiming the effect of our intervention on actual behavior or identity change. However, we use the participants' pre- and post-intervention responses and questionnaire answers as a proxy of behavior change and an indication of the trend of a possible behavior change.

In addition to our main hypotheses, we also explored the effect of ESV on the representation of the ideal self and compared it to other modalities. We hypothesize that ESV makes the ideal self more vivid, similar, and preferable, compared to the text-only and control conditions. We measure this with the adapted Future Self Continuity Questionnaire \cite{sokol2020development}. We also look at participants' free responses to contextualize the quantitative results.


\begin{figure}[b!]
\vspace{-10px}
  \centering
  \includegraphics[width=1\linewidth]{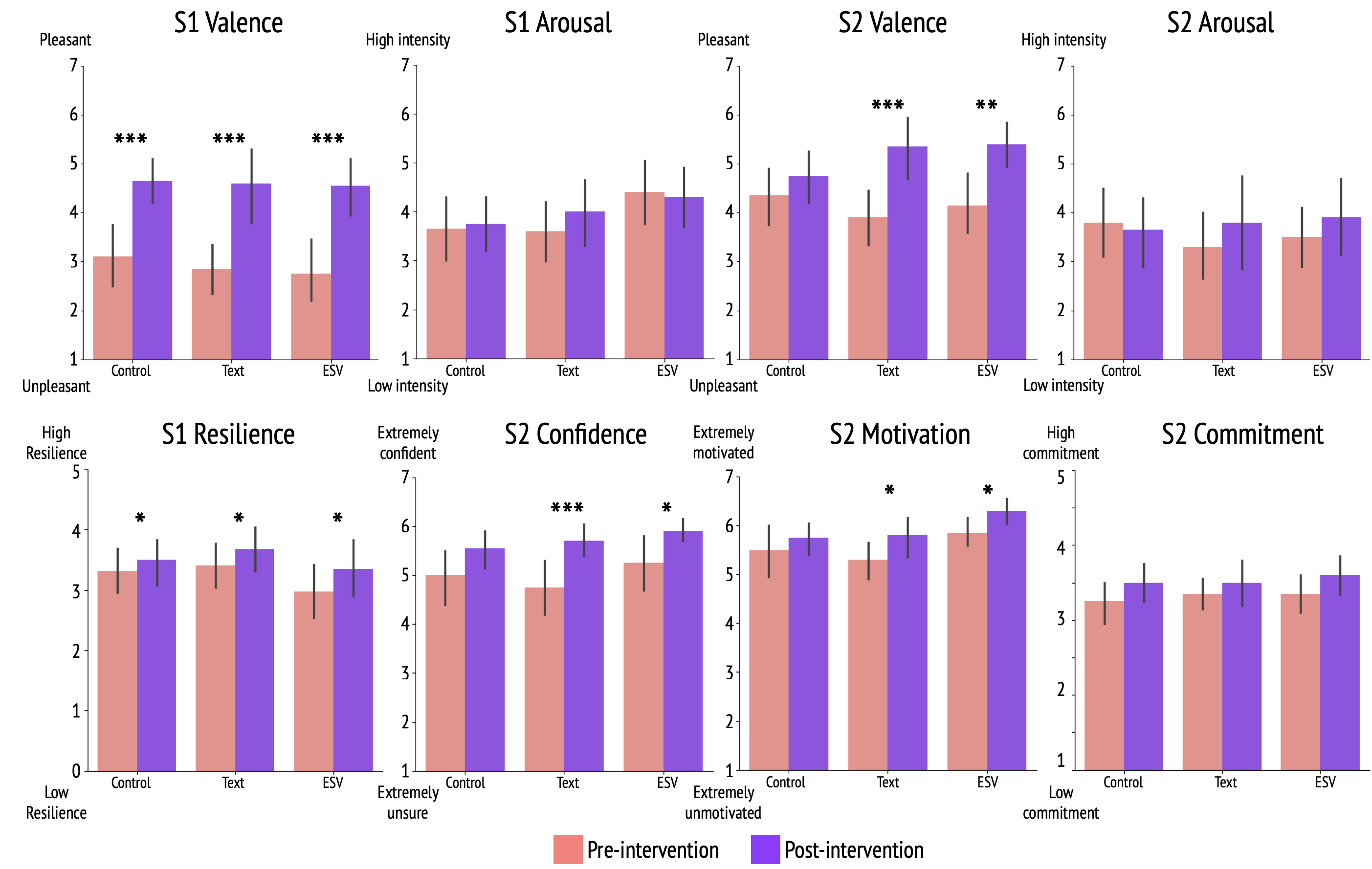}
  \caption{Questionnaire results of the main research outcomes pre- and post- intervention. Error bars: SE. `S1': Scenario 1, `S2': Scenario 1; *: $<$0.05, **:$<$0.01,***:$<$0.001}
  \label{fig:prepost}
\end{figure}

\section{Results and Analysis}
We begin by describing the methods for analyzing the quantitative and qualitative results. We first present the results of our main hypotheses of the effect of the intervention on individuals' affective states and perceptions towards their goals by comparing the results before and after the intervention. Then, we compare results between the conditions to dissect the effect of the individual components of the ESV intervention.

\subsection{Analysis methods}
For the quantitative data, we used a linear mixed effects (LME) model (lme4 package in R \cite{lme4}) to account for the repeated measurement nature of the data, namely each subject has 2 observations: pre-intervention and post-intervention. We used the random intercept model, allowing each subject to have a unique intercept. The predictors of interest are \textbf{Test} (\textit{pre} vs \textit{post}) and \textbf{Condition} (\textit{Control, \textit{Text}, and ESV}), and we control for demographic information like \textit{gender} and age, as well as other aspects related to the intervention like \textit{vividness of mental imagery}, \textit{Alexithymia}, \textit{self-compassion}, and \textit{self-discrepancy}. The main outcomes are \textit{valence}, \textit{arousal}, \textit{resilience}, \textit{confidence}, \textit{motivation}, and \textit{commitment}. Intuitively, these outcomes should not be independent as outcomes from the same person have the same underlying determinants. However, we do not model the correlations among the outcomes at this point. Thus, we fit a random intercept linear mixed effect model for each outcome separately. Note that we encountered collinearity in the condition-related fixed effects for the outcomes \textit{confidence}, \textit{motivation}, and \textit{commitment}. This was due to a near-perfect separation in the pre- and post-intervention data between the correlated conditions. To address this, for these outcomes, we instead conducted a two-way ANOVA on data that was Aligned Rank Transformed \cite{wobbrock2011aligned}, as the scores of the questions did not have a normal distribution (Shapiro-Wilk normality test p<0.001). To identify the levels of our factors that were contributing to observed differences, we conducted a post-hoc pairwise comparison using the emmeans package in R \cite{emmeans}, which produces an adjusted p-value.

We also open-coded and thematically clustered participants’ qualitative questionnaire responses to extract trends. Specifically, we compared participants' reflections on their scenarios after the intervention. The hypothesis is that the intervention might help them reframe the failure or challenges in their pursuit of a goal or a habit. In addition to manual thematic coding, we also used VADER \cite{hutto2014vader} to quantitatively capture the sentiment in the free responses.

\begin{table*}
\begin{tabular}{@{}ccccccc@{}}
\toprule
& \multicolumn{3}{c}{Scenario 1: Overcome a Failure}              & \multicolumn{3}{c}{Scenario 2: Struggle to Persist}             \\ \cmidrule(lr){2-4}\cmidrule(lr){5-7}
& Pre-intervention & Post-intervention & $\Delta$Post-Pre        & Pre-intervention & Post-intervention & $\Delta$Post-Pre        \\ \midrule
\textbf{Control}& -0.170      & 0.346       & 0.516              & 0.206      & 0.318       & 0.113              \\
\textbf{Text}& -0.122      & 0.241       & 0.363              & 0.033      & 0.405       & 0.371              \\
\textbf{ESV}& -0.267      & 0.300       & \textbf{0.567}& 0.128      & 0.587       & \textbf{0.449}\\ \bottomrule
\end{tabular}
\caption{VADER compound scores for pre- and post- affective states. negative: $\leq$-0.05; neutral: $>$-0.05 and $<$0.05; positive: $\geq$ 0.05. Bolded numbers indicate the highest value across conditions.}
\label{tab:vader-emotion}
\vspace{-18px}
\end{table*}

\begin{figure}[b!]
\vspace{-10px}
  \centering
  \includegraphics[width=1\linewidth]{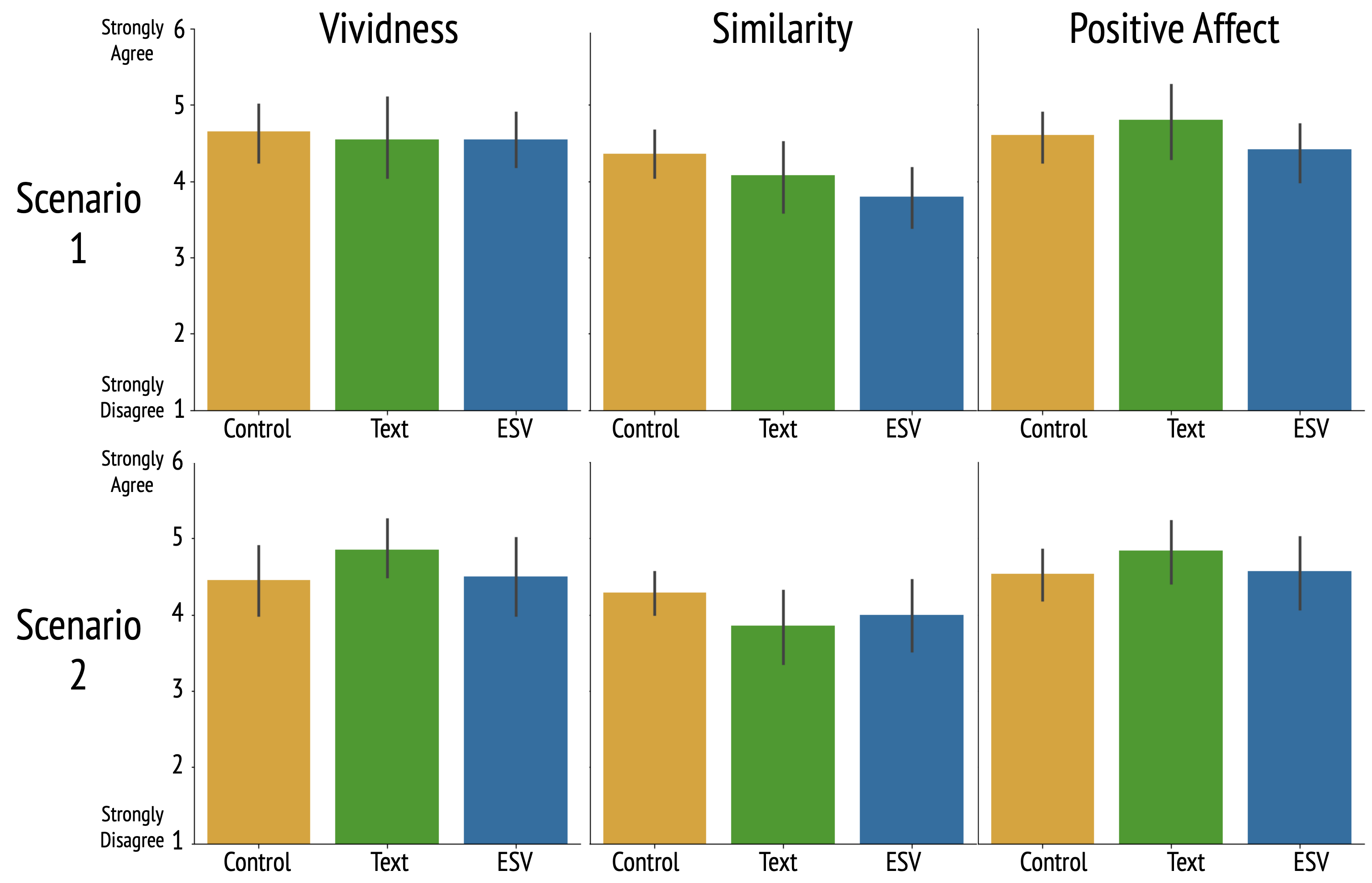}
  \caption{Between-condition comparison of post-intervention ratings of Vividness, Similarity, and Positive Affect towards the Ideal Self. Error bars: SE.}
  \label{fig:fscq}
\end{figure}
 
\subsection{Comparison between pre- and post- intervention}
First, we looked within each condition whether the intervention has any effect on the outcome variables by comparing the responses before and after the intervention.

\subsubsection{Affective states in terms of valence and arousal}
Overall, the valence ratings became significantly (p<.001) more positive across most conditions for both scenarios, but the trends in the change in arousal ratings vary (Figure \ref{fig:prepost} top row). 
Participants overall felt unpleasant upon describing the negative scenarios related to goals pre-intervention and felt more positive after the intervention. Note that across conditions, Scenario 1 ``failure of achieving a goal'' had a more negative pre-intervention valence score compared to Scenario 2 ``difficulty maintaining a habit'' (Scenario 1: $\mu$ = 2.91, $\sigma$ = 0.81; Scenario 2: $\mu$ = 4.16, $\sigma$ = 0.81).
For Scenario 1, the LME model also revealed \textit{self-compassion} and \textit{self-discrepancy} had a significantly negative effect on arousal (self-compassion: $\beta$ = -1.08, p< 0.001; self-discrepancy: $\beta$ = -0.28, p< 0.01). 

We also analyzed participants' free-text explanations about their affective state ratings to contextualize the quantitative data. 

For Scenario 1, the \textit{ESV} condition showed the most noticeable shift from negative to positive emotions: ``\textit{I felt the rejection all over again and imagined what could have been if I had been offered the position}'' (P43, pre-intervention); ``\textit{Made me feel hopeful to better days}'' (P43, post-intervention). The \textit{Control} group had a modest improvement in emotional state, and the \textit{text} group had the least consistent change, with varied responses: ``\textit{Made me feel sad that I didnt achieve that goal. My life would be really different if I had done that}'' (P33, pre-intervention); ``\textit{Made me feel happy about a sad situation. I dont really feel intensity one way or the other}'' (P33, post-intervention).

For Scenario 2, all groups showed some improvement in emotional state.
The \textit{ESV} condition again demonstrated the most noticeable positive shift.
``\textit{The scenario made me feel slightly unpleasant with a pretty high intensity emotional state...}'' (P60, pre-intervention); ``\textit{Made me feel optimistic and hopeful. Feeling more motivated and encouraged}'' (P60, post-intervention).
The \textit{text} condition showed a more consistent positive change in Scenario 2 compared to Scenario 1: ``\textit{When I think about my weight I feel hopeless}'' (P24, pre-intervention); ``\textit{The positive affirmation made me feel like this was worth achieving}'' (P24, post-intervention).
The \textit{Control} condition had the least dramatic change, but still trended positive: ``\textit{Thinking about not being on a daily exercise routine makes me feel angry and disheartening}'' (P4, pre-intervention); ``\textit{motivated me to reflect on my approach and remain determined to find a way to make the change}'' (P4, post-intervention).


The VADER compound scores indicate that for both scenarios, once again, all three conditions show a transition from negative or neutral sentiment to a strong positive sentiment (Table \ref{tab:vader-emotion}). The \textit{ESV} condition appears to be the most effective in improving emotional states across both scenarios.

\subsubsection{Resilience}
For Scenario 1, we looked at participants' resilience towards failure of achieving a goal. Overall, participants across conditions had a significant (p<0.05) increase in resilience after the intervention (Figure \ref{fig:prepost}, bottom left). The LME model revealed that \textit{self-compassion} had a significantly positive effect on resilience ($\beta$ = 0.62, p< 0.001). This shows that the more self-compassion one has, the more resilient one is post-intervention.

We analyzed the themes of participants' post-intervention responses to the question ``what was the reason you didn't succeed?''. While all three conditions predominantly attribute failure to internal factors, the \textit{ESV} condition emphasized personal shortcomings and emotional states: ``\textit{A lack of encouragement, time, and especially self confidence}'' (P46), ``\textit{I wasn't in the right headspace}'' (P50), and ``\textit{mindset and lack of commitment}'' (P60), while \textit{Text} and \textit{Control} conditions focused on external factors: ``\textit{There were outside factors out of my control that contributed to regaining the weight}'' (P5), and ``\textit{The reason I didn't succeed was due to unexpected delays and ineffective time management}'' (P28). On the other hand, the \textit{ESV} condition appeared to have more negative residual feelings and focused on current feelings whereas the \textit{Text} and \textit{Control} conditions focused more on future-oriented thinking. 

The VADER analysis of the free responses shows a similar trend where the \textit{ESV} condition had the least positive sentiment when describing their feeling about the failure post-intervention (Table \ref{tab:vader-scenario} Scenario 1, Feeling). On the other hand, it seems like the \textit{ESV} condition had less negative sentiment when describing failure attribution post-intervention (Table \ref{tab:vader-scenario} Scenario 1, Reason).



\subsubsection{Confidence and Motivation}
For Scenario 2, we looked at participants' confidence and motivation towards keeping up with a personal habit. Overall, participants across conditions had a significant (p < 0.05) increase in both confidence and motivation after the intervention (Figure \ref{fig:prepost}, bottom). It is worth noting that the pre-intervention scores for both outcomes are relatively high (confidence, pre-intervention: $\mu$ = 5.03, $\sigma$ = 0.73; motivation, pre-intervention: $\mu$ = 5.56, $\sigma$ = 0.54; max score = 7), leaving less ``room for improvement''.

Analyzing participants' responses to the question ``How do you feel now about your ability to establish the personal habit?'', the \textit{ESV} group demonstrated more consistent positive feelings and self-belief: ``\textit{I believe in myself}'' (P55), ``\textit{It's hard to do but I am going to keep trying'}' (P58), ``\textit{I feel more capable of taking it on}'' (P41). On the other hand, the \textit{Text} and \textit{Control} groups expressed less certainty: ``\textit{I think it can be achieved}'' (P25), ``\textit{I feel like I need more exercise, but I feel so limited}'' (P5), and more measured optimism and more awareness of the challenge: ``\textit{I feel mildly overwhelmed but cautiously optimistic that I can achieve this goal}'' (P40), ``\textit{I feel optimistic but recognize the challenges}'' (P20). The VADER analysis of the free responses also shows the \textit{ESV} condition had the most positive sentiment when describing their feeling about their capability of habit forming (Table \ref{tab:vader-scenario} Scenario 2, Feeling).

\begin{table*}
\begin{tabular}{@{}ccccccc@{}}
\toprule
& \multicolumn{2}{c}{Scenario 1: Overcome a Failure} & \multicolumn{2}{c}{Scenario 2: Struggle to Persist} \\ \cmidrule(lr){2-3}\cmidrule(lr){4-5}
& Reason & Feeling & Reason & Feeling \\ \midrule
\textbf{Control}& -0.019& \textbf{0.128}& -0.066& 0.380\\ 
\textbf{Text}& -0.103& 0.089& \textbf{0.102}& 0.273\\ 
\textbf{ESV}& \textbf{0.046}& 0.011& -0.042& \textbf{0.400}\\ 
\bottomrule
\end{tabular}
\caption{VADER compound scores for perceived reasoning of and feeling towards the challenge. negative: $\leq$-0.05; neutral: $>$-0.05 and $<$0.05; positive: $\geq$ 0.05. Bolded numbers indicate the highest value across conditions.}
\label{tab:vader-scenario}
\vspace{-10px}
\end{table*}

\subsubsection{Commitment}
For Scenario 2, we also looked at participants' commitment towards their goal (establishing a personal habit). Overall, we observed an increase, albeit not significant, in the goal commitment scores post-intervention (Figure \ref{fig:prepost}, bottom right).
We again reviewed participants' post-intervention free-response to questions about the reason for the difficulty behind establishing their personal habit, and how they feel now about their ability to establish the personal habit. Overall, both the \textit{ESV} and \textit{Text} conditions focused more on the self as the attribute of the difficulty: ``\textit{Self discipline and a lack of affirmation on my small successes}'' (P58), ``\textit{Myself. I just don't have the right positivity towards myself}'' (P34). In turn, they emphasized that they have more agency over the ability to obtain the habit: ``\textit{But the reasons feel a little less important and the reality that if I try I can do anything I want despite the difficulties}'' (P25), and they just needed to be reminded about the possibility of success: ``\textit{I forget at times that the personal habit is achievable}'' (P44). In contrast, the \textit{Control} condition showed less enthusiasm and focus on external factors as the cause for the difficulty: ``\textit{i guess I dont care about it as much as I want to}'' (P10), ``\textit{It is difficult to form new life style choices to change habits}'' (P6), ``\textit{So many other responsibilities that feel more pressing in the moment}'' (P14).

\subsubsection{Clustering of Responders and Non-Responders}
In addition to the main hypotheses and aforementioned results from the LME model, we tried to understand the difference between responders and non-responders to this type of cognitive intervention. We performed a hierarchical clustering of participants of all conditions. We found that for scenario 1, people who were the least respondent to the intervention were older and had higher self-compassion, whereas the most respondent group on average had a noticeably younger age, lower self-compassion, and a higher self-discrepancy. This makes sense intuitively, as people who are already self-compassionate might not need the intervention. For scenario 2, we found that people who had higher alexithymia did not respond to the intervention, which could be that they were less susceptible to the ``self pep-talk''. The additional results are provided in the Appendix \ref{sec:clustering}.



\subsection{Comparing Effectiveness of Ideal Self Representation Between Conditions}

We measured the distance between the individual's current self and their ideal self using the adapted version of the Future-Self Continuity Questionnaire, where we replaced ``future-self'' with ``ideal-self''. Overall, the post-intervention ratings of the three aspects: vividness, similarity, and positive affect were above neutral (score = 3.5) (Figure \ref{fig:fscq}). 

We did not observe any significant differences between the conditions in the quantitative results (Figure \ref{fig:fscq}), but we found more nuanced differences in the qualitative responses. The \textit{Control} condition group expressed a positive yet low intensity attitude towards the ideal self concept and felt a sense of distance: ``\textit{I feel happy thinking about my 'ideal self' as it's something that everyone hopes to achieve. I don't really feel that strongly about it since it seems distant to me}'' (P2), ``\textit{The imagining did not invoke too much internal emotion}'' (P14). The \textit{Text} condition group responded more strongly towards the generated ideal-self responses and appreciated the affirmation from a first-person perspective: ``\textit{I really liked how the AI spoke in first person. It helped me feel like I was saying in affirmation to myself and that I could actually achieve what was written}'' (P26). Similarly, the \textit{ESV} condition felt a strong sense of positivity and empathy from a first-person's perspective: ``\textit{It was encouraging to hear a generated response using my voice to hear my goals along with adjectives describing myself}'' (P56). ``\textit{Its like hearing myself speak positively to me helps me be more positive}'' (P55). The generated ideal-self voice activated individuals: ``\textit{...It was exactly what I needed to hear}'' (P59), ``\textit{It made me enthusiastic enough to come up with a plan on how to undertake the problem and generate a working solution}'' (P53). More concretely, participants in the \textit{ESV} condition explicitly mentioned being able to more vividly visualize an ideal version of themselves: ``\textit{made me visualize a version of myself that is a bit more positive}'' (P48), ``\textit{I have a positive opinion of the voice and text from the scenario-in fact, can even envision her in my mind (she doesn't look like me, doesn't have the same job as me...) but I like how articulate `I' was and would feel good about myself saying what `she' said}'' (P52).





\section{System Characterizations}

The goal of the characterizations is to confirm our pipeline's ability to generate the intended emotion. The output of our pipeline consists of text and speech output, and both contribute to the overall sentiment and attitude of the text. Complementary to the user study results which provide a subjective, qualitative evaluation of the emotional quality of the output, the technical validation tests are meant to provide a quantitative, objective rating of the emotional quality of the generation.


\begin{table}[b!]
\vspace{-15px}
\begin{tabular}{@{}ccccccc@{}}
\toprule
&\textbf{Scenario 1}& \textbf{Scenario 2} \\ \midrule
& \textit{Mean (SD)} & \textit{Mean (SD)} \\
\textbf{Text}&0.57 (0.59)& 0.80 (0.33)\\ 
\textbf{ESV}& 0.65 (0.44)& 0.78 (0.18)\\ 
\bottomrule
\end{tabular}
\caption{VADER results of the generated texts. negative: $\leq$-0.05; neutral: $>$-0.05 and $<$0.05; positive: $\geq$ 0.05. }
\label{tab:vader-gen-text}

\end{table}

\subsection{Validating the sentiment of the text content}

We again chose VADER \cite{hutto2014vader} to quantitatively analyze the sentiment of the generated texts. Overall, the results show that generated ideal-self text responses were positive (Table \ref{tab:vader-gen-text}), and more positive for Scenario 2 than Scenario 1, likely due to the nature of the context of the Scenario 2 focusing on future habits rather than past failures.

We compared the words participants used to describe the ideal self against their perceived attitude of the generated text. In both scenarios, the generated text seems to have been perceived as more overtly positive and emotionally supportive than the participant-provided ideal self characteristics. The ideal self words tend to be more focused on personal qualities and actions, while the generated texts often reflect the emotional impact of the generated content. In addition, the ideal self words are more action-oriented (e.g., ``disciplined'', ``consistent'', ``organized''), while the text descriptions focus more on emotional states and motivation. Some key ideal self characteristics like ``disciplined'' and ``hardworking'' do not appear in the text descriptions, suggesting the generated text might not have emphasized these aspects as much. There is a greater alignment between the ideal self and the generated text descriptions in Scenario 2. In both cases, the generated text seems to have successfully captured many aspects of what people aspire to be, particularly in terms of positivity, resilience, and motivation.

Participants' qualitative responses also show the generated text contents brought a positive perspective to the negative scenarios: ``\textit{Made me feel happy about a sad situation}'' (P33), ``\textit{It was a little too positive for me but the intensity wasn't overwhelming}'' (P38), ``\textit{I really felt like I did learn from the experience in a positive way}'' (P24). In addition, the actual content was perceived as empathetic and realistic: ``\textit{I feel like a positive message was delivered that showed care and understanding} ''(P45) ``\textit{...because it understood what i was conveying and knew that i would have bounce back from any circumstances}'' (P23), ``\textit{I felt like the response was something I would actually think. It was uplifting but realistic...}'' (P26).

\begin{table}[t!]
\begin{tabular}{@{}ccccccc@{}}
\toprule
&\textbf{Ratings} \\ \midrule
& \textit{Mean (SD)} \\
  \textbf{Familiarity}&3.45 (0.94)\\
 \textbf{Similarity}& 2.90 (0.90)\\
 \textbf{Naturalness}& 3.05 (1.32)\\
 \textbf{Quality}& 3.90 (1.02)\\
\bottomrule
\end{tabular}
\caption{Participants' ratings of various aspects of the ESV speech on a scale of 1 to 5 where a higher rating is better. }
\label{tab:clone-quality}
\vspace{-15px}
\end{table}

\subsection{Validating the audio quality of the speech}

For the \textit{ESV} condition, we asked participants questions about hearing their own voice and how they would rate the generated self-voice. These participants had limited to no experience with voice cloning prior to the study ($\mu$ = 1.15, $\sigma$ = 0.49, 1=`Once', 2=`A few times a year'). On average, they had a neutral to slightly negative feeling towards listening to their own voice ($\mu$ = 2.85, $\sigma$ = 1.31, 2=`Slightly no', 3=`Indifferent'). Overall, these participants rated the generated self-voices to be familiar, similar to their own, natural, and have good quality (Table \ref{tab:clone-quality}). In contrast to reactions to listening to their true voice, overall no participants responded negatively to the generated self-voice. Most reported being pleasantly surprised by the resemblance of the voice: ``\textit{Caught off guard with how striking the resemblance is}'' (P54), ``\textit{I couldn't believe how quickly it cloned my voice and how accurate it was at reflecting my tone. Made me excited for the future}'' (P55), ``\textit{Really got close to what I actually sound like}'' (P54).

\begin{table}[t!]
\begin{tabular}{@{}ccccccc@{}}
\toprule
&\textbf{Scenario 1}& \textbf{Scenario 2} \\ \midrule
& \textit{Mean (SD)} & \textit{Mean (SD)} \\
\textbf{Angry}&0.0020 (0.0049)& 0.0060 (0.022)\\ 
\textbf{Happy}& 0.52 (0.46)& 0.76 (0.38)\\ 
\textbf{Neutral}& 0.31 (0.40)& 0.22 (0.37)\\ 
\textbf{Sad}& 0.061 (0.22)& 0.0051 (0.013)\\ 
\textbf{Unknown}& 0.10 (0.22)& 0.011 (0.032)\\ 
\bottomrule
\end{tabular}
\caption{Summary results of emotion classification (emotion2vec) of the generated ESV speeches}
\label{tab:emo2vec}
\vspace{-15px}
\end{table}

\begin{figure}[b!]
\vspace{-10px}
  \centering
  \includegraphics[width=1\linewidth]{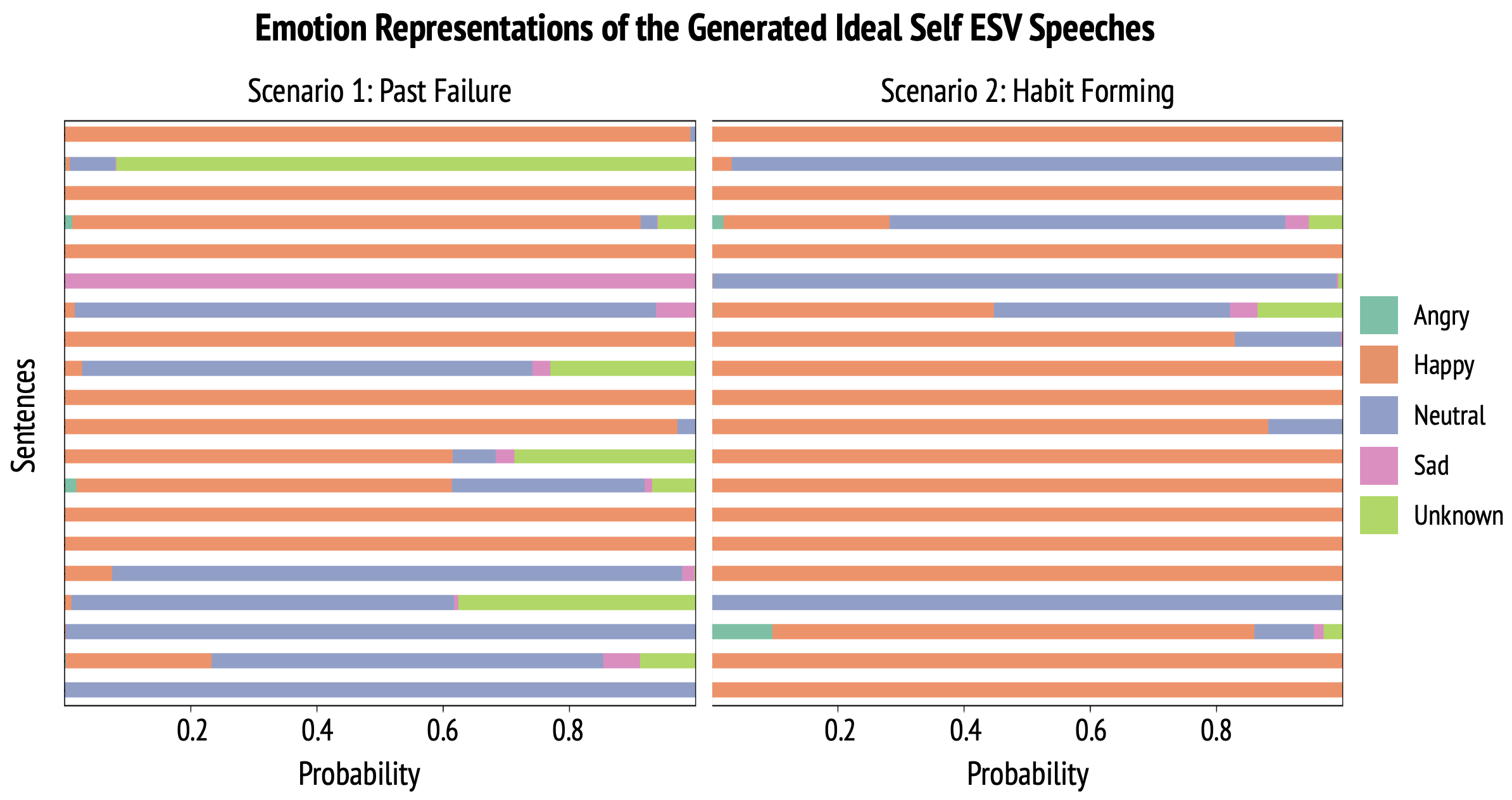}
  \caption{Emotion classification results (emotion2vec) of the emotions represented in the generated ESV speeches.}
  \label{fig:emotion2vec}
  
\end{figure}

\begin{table*}

\begin{tabular}{@{}ccccccc@{}}
\toprule
& \multicolumn{2}{c}{Scenario 1: Overcome a Failure} & \multicolumn{2}{c}{Scenario 2: Struggle to Persist} \\ \cmidrule(lr){2-3}\cmidrule(lr){4-5}
& Positivity & Emotional Expressiveness & Positivity & Emotional Expressiveness \\ \midrule
& \textit{Mean (SD)} & \textit{Mean (SD)} & \textit{Mean (SD)} & \textit{Mean (SD)} \\
\textbf{Text}& 0.58 (2.09)&0.95 (1.58)& 1.47 (1.17)& 0.21 (1.72)\\ 
\textbf{ESV}& 0.94 (1.98)& 0.94 (1.73)& 1.22 (1.96) & 0.89 (1.81)\\ 
\bottomrule
\end{tabular}
\caption{Summary of participants' parameter settings. The range of the values are -3 to +3.}
\label{tab:parameters}
\vspace{-10px}
\end{table*}

\subsection{Validating the emotional quality of the speech}

Once again, we compared the words participants used to describe their ideal self characteristics against their perceived attitudes of the generated audio. In both scenarios, the generated voice seemed to have been perceived as more emotionally nuanced and expressive compared to the more action-oriented ideal self characteristics.
The voice appeared to have successfully conveyed a sense of positivity and confidence that aligns well with people's ideal self-perceptions. The realism and accuracy of the voice were frequently mentioned, suggesting that it felt authentic to listeners despite being generated. Some respondents found the voice to lack emotion or sound monotone, indicating that the emotional expressiveness might not have been consistent for all listeners. The voice descriptions included words like ``calm'' and ``balanced'', which did not directly appear in the ideal self words but may reflect a desired state of being focused and organized. The voice was often described as ``expressive'', which seemed at times to be perceived as overtly optimistic.

We chose the state-of-the-art emotion2vec model \cite{ma2023emotion2vec} to analyze the sentiment of the speech. We processed the generated speeches for the \textit{ESV} conditions and we analyzed them at a sentence level. Figure \ref{fig:emotion2vec} and Table \ref{tab:emo2vec} show the emotion2vec results. Overall we see that the emotional representations of the two scenarios are distinct: with sentences in Scenario 2 sounding mostly ``happy'' while ones in Scenario 1 sounding ``happy'' but more ``neutral''. 

Participants reported feeling more positive as a result of hearing affirmations spoken with their voices: ``\textit{I liked hearing `my' voice with positivity and determination}'' (P47), ``\textit{I believe the generated audio would make for a great affirmation to remind myself that the goal is achievable}'' (P44), ``\textit{It made me feel slightly better, because the voice was saying to believe in myself. Honestly, I should!}'' (P58).



\subsection{Parameterization}

Table \ref{tab:parameters} shows a summary of the participant-tuned parameters of the generated text and voice. Overall, the mean values are all above 0 (on a scale of -3 to +3), indicating that on average people prefer more positivity and expressiveness in the generated responses. However, these values have significant variance, pointing to the nuanced differences among individual preferences.
We then contextualized the chosen parameter values by analyzing participants' free responses to explaining their chosen parameters. Many participants chose parameters that aligned with their emotional state or how they wanted to feel about the situation. Some aimed for positivity to make themselves feel better, while others chose more negative sentiments to match their experience. Many participants in both conditions tried to find a balance between different aspects (e.g., positivity vs. realism, expressivity vs. calmness). For Scenario 2, participants were generally more focused on creating positive, motivational outputs, likely due to the forward-looking nature of forming a new habit. The parameters were often chosen with the specific goal of habit formation in mind, aiming for a tone that would be most conducive to achieving the goal.

\section{Discussion}

We begin by discussing immediate findings based on the results of the study, and then we discuss the broader implications for theories related to ideal-self and goal-oriented behavior change and implications for designing future intervention systems.

\subsection{Summary of Findings}

Our study on the Emotional Self Voice intervention reveals a nuanced picture of its effectiveness in addressing negative aspects of goal achievement. While we observed positive trends across all conditions, the \textit{ESV} condition showed particular promise in certain areas.

The intervention generally succeeded in inducing more positive affect across scenarios, partially supporting H1 and H3, and aligning with prior work showing positive effect for emotional synthetic voices~\cite{cohn2020differences,nass2001effects}. However, the differences between ESV and other conditions were not as pronounced as initially hypothesized. This suggests that while ESV can be an effective tool, the act of reflection itself, regardless of modality, may have inherent benefits. Regarding resilience (H2), and confidence and motivation (H4), we observed improvements across conditions. Goal commitment (H5) showed a positive trend but didn't reach statistical significance. This could be due to the short-term nature of our study, as commitment may require more time to solidify.

We compared the scenarios participants provided to the two prompts. While participants mentioned similar goals for both scenarios, such as health and fitness, the two sets of goals differ fundamentally in their scope and emotional landscape. The first set uncovers profound psychological struggles around significant life milestones, career advancement, and personal transformation, whereas the second set reveals tactical, routine-level obstacles centered on personal wellness. Essentially, the first represents high-stakes personal and professional aspirations, and the second represents incremental personal growth. Correspondingly, the contrasting nature of our two scenarios revealed interesting differences in intervention effectiveness. Scenario 1 elicited more mixed emotional responses, with some participants experiencing negative feelings from revisiting past failures: ``\textit{I just feel worse having spent time thinking about it}'' (P41), ``\textit{Felt triggered a bit and somewhat unpleasant}'' (P21).
 This highlights the delicate nature of addressing past shortcomings and the need for careful framing in interventions. We also observed self-compassion as a strong predictor for resilience, which the ESV system could leverage. Scenario 2, focused on future habits, more consistently yielded positive responses. This forward-looking perspective seemed to align well with the ideal-self concept, possibly because it allowed participants to envision improvement without the emotional baggage of past failures. These differences underscore the importance of context in behavioral interventions.

The \textit{ESV} condition, while not always significantly outperforming others, showed unique benefits. \textit{Text} and \textit{Control} conditions demonstrated the power of reflection itself. The \textit{ESV} condition added an extra dimension of engagement and personalization. The self voice shifted more attention onto oneself in terms of the attributions of the reason for the failure and challenges. Some liked to think of themselves as similar to their ideal self: ``\textit{I am pleasant with my ideal self which is why I feel I am similar to my ideal self}''(P4), and a stronger alignment between the actual and ideal self brings more optimism, but the opposite is true as well: ``\textit{Thinking of my ideal self made me think of my flaws and how difficult it would be to become my ideal self}'' (P6).


Our technical characterization revealed that the generated text and voice content was consistently positive, especially for Scenario 2. The above-average ratings for voice familiarity, similarity, and quality in the \textit{ESV} condition suggest that our voice cloning technology achieved its goal of creating a believable ``ideal self'' voice. The parameterization findings offer intriguing insights into user preferences for idealized self-representations. The general preference for positivity and expressiveness, balanced with realism, suggests a sweet spot for effective interventions.

\subsection{Implication for theory}
We contribute to the literature on self-discrepancy. The effectiveness of the ideal self voice in improving affect and motivation supports the idea that reducing the perceived gap between actual and ideal selves can lead to positive outcomes~\cite{higgins1994ideal}. Our investigation highlights the potential of self-voice as a modality for these types of interventions. More specifically, we contribute findings on the novel use of generative models for self-voice as an intervention. The unique benefits of the \textit{ESV} condition suggest that hearing one's own voice may create a more immersive and impactful experience than text alone.

The positive reception of the idealized self-voice, despite many people's typical aversion to hearing their own voice~\cite{holmes2018familiar,kim2012visualizing}, suggests that individuals may internalize the characteristics of the ``improved'' voice they hear. Many previous works have investigated the negative aspect of self-voice~\cite{daryadar2015effect,holzman1966voice,holzman1966listening,kim2024myvoice}; we show how generative self-voice, instead of negatively affecting one's emotion and behavior, can bring positive influences. 

The differential effects between past-focused and future-focused scenarios highlight the importance of framing in goal-related interventions, potentially extending our understanding of how the temporal and cognitive focus impacts goal pursuit. For example, the idea of a past failure brings out unpleasant feelings and conflates with ongoing goals. However, reframing past failure as a part of an ongoing goal might point the individual toward a more positive outlook. In general, designers can look into the psychology and behavioral science literature and test out different reflection prompts with the ESV system.


\subsection{Implications for designing generative self-voice systems for behavioral intervention}


Through the study, we also attempted to characterize people's preferences for the generated ideal-self responses in terms of content, speech characteristics, and attitude. The parameterization step also allows us to understand a bit deeper about the individual preferences on these parameters. The preference for positive yet realistic content suggests that interventions should strive for an optimistic tone while maintaining credibility. 

There were also a couple of design choices that we have made; although they were not explicitly tested, we highlight here how they may have contributed to a successful intervention evidenced by the positive effects on the participants in the \textit{Text} and the \textit{ESV} conditions. First, we wanted to achieve a balance between self-reflection and assistance from the generative tools. We used a realistic scenario provided by the participants, which elicits genuine emotion rather than a synthetic task. Second, instead of providing one possible generation, we provide two parameters on a slider to give people more fine-grained control over the generated response. 

Future work should look into the effect of people's prior experience with or attitudes towards AI and voice technologies. Prior work around AI chatbot therapist have found mixed results. Some showed positive attitude towards mental health chatbots \cite{abd2021perceptions}, while others found the attitudes varied by demographics\cite{benda2024patient}. Compared to human-generated responses, AI-generated responses have been rated to have better perspective-taking and empathic concern \cite{yonatan2024comparing} while revealing nuanced differences in the focus and scope of the content \cite{li2024skill}.
Different from AI chatbot-based interventions, the role of AI in our system is not as salient given the responses were generated from a first-person perspective, and the awareness of AI as the source can alter the perception of the provided responses \cite{jain2024revealing}. We further see how the similar-attraction effect \cite{reeves1996media} plays a role in hearing the generated response in the user's own voice.

Although the interaction presented in the prototypical system is time-bounded (i.e., limited by what the study allows) and we cannot claim the longer-term effect on behavior change and habit formation, we believe that our system's ability to generate emotional self-voice that can adapt to the individual's scenario and ideal self can further adapt to their ever-changing needs in a longitudinal, real-world setting.




\section{Ethical Considerations}

This technology could pose privacy issues. However, this is not unique to this system. Any system that uses one's information to generate hyper-realistic versions can pose an issue of impersonation and identity threat. Active research is being done on identifying deepfakes in the forms of image, audio, and video \cite{chen2020generalization,zhou2021joint,rana2022deepfake,groh2024human}.

In addition, the models we used in the system are not ``white-boxed''. Our proof-of-concept system prioritizes the speed and stability of the generation process, and thus we opted for third-party, cloud-based models. It would be better if the models were open-sourced and locally hosted. The latter would also further improve the privacy of the system.

While in this work we have shown the positive benefits of exposing one to a generated ideal-self, it is not hard to imagine how this technology can be used as a form of unpleasant marketing strategy or worse, malicious manipulations that capitalize on one's emotion and desire to better themselves. Specifically, self-voice implies the intervention is experientially personalized to an individual, which could make them more susceptible to manipulation. Thus, we cautioned the participants from blindly taking action based on the AI-generated responses, and we need to understand the long-term effect of using this intervention consistently.

\section{Limitations \& Future Work}

\textit{Uncanny Valley of Generated Self Voice}.
Prior work has shown that an increase in voice familiarity can lead to eerie feelings due to uncanny valley \cite{chan2021kinvoices,holzman1966voice,kimura2018auditory}. Although not mentioned explicitly by the participants, we acknowledge the unpleasantness that comes with uncanny valley. The fact that people did not find the generated voices disturbing could be due to knowing that it was an AI-generated version of their own voice \cite{hughes2013like}. 

\textit{Limitations of Self Voice Replication and Range of Emotional Expressiveness.}
The quality of the generation is affected by accents.
We screened for people who speak fluent English, but this does not explicitly exclude accents. Although no participants explicitly mentioned the dissimilarity of the accent of the generated voice, in our early experiments, we did notice having a non-American accent reduces the naturalness of the generated voice. Thus, the findings of this study might not generalize to cases where individuals' accents are not well captured in the generated ESV. The current system has good emotional quality but the range of expression is limited to how the model was trained and the capabilities exposed to developers. The emotional range of the current implementation depends on the emotional TTS model's interpretations of the emotion of the input text and the expressiveness of the samples provided for voice cloning. The range of emotional tonalities could be expanded and the preciseness of the expression, for example, being able to express different emotional attitudes of the same text \cite{wu2024laugh}. Individuals can then experiment with hearing their own voice speaking the same text but with different attitudes.

\textit{Implication on Long-term Behavior Change and Habituation}.
As an initial step, our study investigated how ESV affects one's perception of an event, which can be a proxy of the direction of behavior change. Future work should investigate the efficacy of this technique on behavior change with a longitudinal study with actual behavioral tasks. 

A follow-up study should explore the potential psychological impact of repeated exposure to one’s ideal self-voice, for example, changes in the user's self-identity over time, whether the intervention promotes the internalization of ideal-self traits or cognitive dissonance.

In addition, we did not study the habituation effect towards the intervention, where users may become desensitized to hearing their own voice over time. While the study of habituation is beyond the scope of the study, we speculate the self-relevant stimuli, i.e., cloned self voice, have less habituation effect compared to other stimuli such as other voices \cite{rogers1977self}. On the other hand, the content might lead to habituation effect if the same cognitive reframing strategies are used repetitively. To mitigate this, the system could introduce tone, phrasing, or style variability to maintain engagement.

\textit{Ideal-Self within Intrapersonal vs Interpersonal Contexts.} 
We focused on scenarios in the self-development context. In reality, the concept of ideal self also applies to interpersonal communications (e.g., conflict resolution, asking for a raise). We tested the ESV on the interpersonal scenario for our pilot study, and we found that the generated responses of ESV varied more as the interpersonal challenges depended on the opponent's personality and the relationship (positive or negative) between the user and the opponent. Future work should investigate the application of ESV in interpersonal contexts, where the ideal-self concept might adapt to others and the environment \cite{wille2018yourself}.

\textit{Other Voice, Selves and Modalities.}
We focused on the effect of the ideal self in this work. A natural extension of this work is to investigate the effect of an ``ideal-other'', which is essentially a role model or someone else influential. Prior work has compared the effect of self-voice and other-voice, but the voice of an ``ideal-other'' adds another interesting dimension. In this work, we only explored the effect of experiencing the ideal self. It would be interesting to explore the effect of experiencing other versions of the self, such as the current self or the ``ought self'', which is who one or others think one should be \cite{higgins2013patterns}, which might be interesting in interpersonal contexts. Alternatively, exposing one to their current self might evoke negative emotions, but for some, negative emotions might drive more motivation and activation towards a goal. Future systems might have an adaptive strategy of motivating individuals depending on what they need in the moment. Future work can explore the combination of emotional self voice with other modalities, such as seeing an ideal version of the self or embodying an ideal self in an immersive environment, to increase the overall vividness of an ideal self.

\section{Conclusion}

We introduce Emotional Self Voice (ESV), a proof-of-concept system that generates responses of an idealized self, delivered through a personalized, emotionally expressive cloned voice. Focusing on two critical aspects of goal achievement---overcoming failure and struggles with persistence---we demonstrate ESV's potential to shift self-perception and behavior. Our study reveals that engaging with an ideal-self perspective generally increases positive affect and enhances key behavioral drivers such as resilience, confidence, motivation, and commitment across all conditions. While ESV did not consistently outperform other conditions that only involved mental imagination or text-based interventions, it demonstrated unique benefits, particularly in scenarios involving future habit formation. Participants showed high engagement with the ESV technology, expressing surprise and excitement at hearing their cloned voices articulate positive, motivational messages. Future research should explore the long-term effects of ESV, its applications across various behavioral domains, and the ethical implications of using AI-generated self-voices in psychological interventions. The significance of this work lies in its novel integration of advanced AI technologies with psychological theories of self-perception and behavior change. We envision ESV as a ubiquitous, adaptive tool in everyday life, providing personalized support at critical moments of decision-making and self-reflection.


\begin{acks}
The authors would like to thank A. Kapur and M. Cherep for their feedback on the manuscript, J. Shen for creating the teaser figure illustrations, and the study participants for their time.
\end{acks}

\bibliographystyle{ACM-Reference-Format}
\bibliography{references}

\appendix

\section{System Implementation}

\subsection{Prompt for text generation} \label{sec:appendix-text-prompt}

\textbf{Scenario 1:} You embody a specific type of person with given personalities as part of a dramatic theatrical play.  Your task is to imagine yourself as the person with these trait personalities would say to themselves in the given scenario that would help them reframe the situation and be resilient in face of a failure.
You will also be given some settings to refine the emotional affect of sentence: positivity, emotional expressivity. The default value is 0 and the range is -3 to +3

Requirements:
Your response must refer to what happened in the scenario.
You must express the attitudes and emotions saliently. 
You can add vocal bursts, natural vocal inflections and discourse markers.
Never use markdowns or emojis. Use first-person.
Keep the response short within two sentences.

\textbf{Scenario 2:} 
You embody a specific type of person with given personalities as part of a dramatic theatrical play.
Your task is to imagine yourself as the person with these trait personalities would say to themselves in the given scenario that would encourage them to keep up with the habit. You should try to embody the person when their habit has become their identity. Use the template of ``I am a xxx person.'' You will also be given some settings to *finetune* the emotional affect of sentence: positivity, emotional expressivity. The default value is 0 and the range is -3 to +3. The personalities have more priority than the settings.

Requirements:
You must express the attitudes and emotions saliently. 
You can add vocal bursts, natural vocal inflections and discourse markers.
Never use markdowns or emojis. Use first-person.
Keep the response short within two sentences.

\subsection{Text for voice cloning} \label{sec:appendix-cloning}

**Read the following sentences with a happy tone:**
\begin{itemize}
    \item ``Hey there! It's such a beautiful day, isn't it? The sun is shining, and I just feel so alive!''
    \item ``I just got great news! I got the job I've always wanted. I can't stop smiling!''
    \item ``Wow, this cake tastes amazing! You have to try it. It’s the best I’ve ever had.''
    \item ``Guess what? We’re going to the beach this weekend! I can’t wait to splash in the waves.''
\end{itemize}
**Read the following sentences with a sad tone:**
\begin{itemize}
    \item ``I can't believe it. My pet ran away. I’ve looked everywhere, but I can’t find him.''
    \item ``It's just one of those days where everything feels heavy. I wish things would get better.''
    \item ``I got the call this morning. I didn’t get the part in the play. I’ve been crying all day.''
    \item ``Why does it feel like every time I try my best, it’s never enough? It’s really hard to stay positive.''
\end{itemize}

\newpage
\section{User study Questionnaire} \label{sec:appendix-survey}
 Here we provide all the questionnaires used in the study.
\subsection{Demographics}
\begin{itemize}
    \item Age
    \item Gender
    \item Perth Alexithymia Questionnaire (PAQ-S) \cite{preece2023perth}
    \item Plymouth Sensory Imagery Questionnaire Sound and Feeling subscale (Psi-Q) \cite{andrade2014assessing}
    \item (\textit{ESV} condition only) Prior Experience with Voice Cloning Technology: Have you ever used voice cloning technology? (Not at all, Once, A few times a year, A few times a month, A few times a week)
    \item State Self Compassion Scale (SSCS-S) \cite{neff2021development}
    \item Big Five personality (ideal and current self) (Mini-IPIP) \cite{donnellan2006mini} 
\end{itemize}

\subsection{Voice Cloning (\textit{ESV} condition only)}
\begin{itemize}
    \item Do you like listening to your own voice? (Absolutely no, Slightly no, Indifferent, Slightly yes, Absolutely yes)
    \item Please explain your responses to the above question
    \item How familiar was the voice? (Very familiar, Somewhat unfamiliar, Neither familiar nor unfamiliar, Somewhat familiar, Very familiar)
    \item How similar was the voice delivering the response to your own voice in terms of: Accent, Speaking Rate? (Not at all similar, Slightly similar, Moderately similar, Very similar, Extremely similar)
    \item How would you rate the naturalness of the audio? (Very unnatural, Somewhat unnatural, Neither natural nor unnatural, Somewhat natural, Very natural)
    \item How would you rate the quality of the audio? (Bad, Poor, Fair, Good, Excellent)
\end{itemize}

\subsection{Pre-intervention}
\begin{itemize}
    \item Move the slider to rate how the scenario made you feel. How pleasant is your emotional state as a result of the scenario? (1---unpleasant, 7---pleasant)
    \item Move the slider to rate how the scenario made you feel. What is the intensity of your emotional state as a result of the scenario? (1---low intensity, 7---high intensity)
    \item Please explain your responses to the above slider questions.
    \item (Scenario 1 only) Resilience: Benchmark Resilience Tool (BRT) \cite{smith2008brief} 
    \item (Scenario 2 only) Motivation: How motivated are you to achieve this goal? (Extremely unmotivated, Unmotivated, Slightly unmotivated, Indifferent, Slightly motivated, Motivated, Extremely motivated)
    \item (Scenario 2 only) Confidence: How confident are you that you will be able to achieve this goal? (Extremely unsure, Unsure, Slightly unsure, Indifferent, Slightly confident, Confident, Extremely confident)
    \item (Scenario 2 only) Goal Commitment Measure scale \cite{hollenbeck1989investigation}
\end{itemize}

\begin{table*}
\begin{tabular}{@{}ccccccccccc@{}}
\toprule
& \textbf{Alexithymia} & \textbf{PSI audio} & \textbf{PSI feeling}   & \textbf{Compassion} & \textbf{Self discrepancy}                         \\ \midrule
&\textit{Mean (SD)}   & \textit{Mean (SD)}   &\textit{ Mean (SD)} & \textit{Mean (SD)} & \textit{Mean (SD)}           \\ 
\textbf{Control} &4.00 (1.33)	&8.53 (1.47)	&8.22 (1.55)	&3.13 (0.54)	&3.40 (1.63)	                             \\
\textbf{Text}&3.99 (1.47) &8.83 (1.24)	&8.30 (2.06)	&3.07 (0.51)	&3.85 (2.29)                               \\
\textbf{ESV}&4.31 (1.17) &8.75 (1.26)	&7.53 (2.62)	&2.80 (0.79)	&5.30 (2.50)
\\ 
\bottomrule
\end{tabular}

\caption{Demographics and Control Variables by Condition.}
\label{tab:demographics}
\end{table*}

\subsection{Post-intervention}
\begin{itemize}
    \item (\textit{ESV} condition only) How would you describe the emotional attitudes of the voice?
    \item How would you describe the emotional attitudes of the text response?
    \item Please explain your choice of generation parameters (Positivity, Expressivity).
    \item Modified Similarity subscale (Q1-4) of Future Self Continuity Questionnaire\cite{sokol2020development} replace ``what you will be like 10 years from now'' with ``your ideal self in this context and ``beliefs'' with ``emotional attitudes''.
    \item Modified Vividness subscale (Q5) of Future Self Continuity Questionnaire\cite{sokol2020development} replace ``what you will be like 10 years from now'' with ``your ideal self in this context.
    \item Modified Positive Affect subscale (Q8-10) of Future Self Continuity Questionnaire\cite{sokol2020development} replace ``what you will be like 10 years from now'' with ``your ideal self in this context and added a question on ``emotional attitudes''.
    \item Move the slider to rate how the scenario made you feel. How pleasant is your emotional state as a result of the scenario? (1---unpleasant, 7---pleasant)
    \item Move the slider to rate how the scenario made you feel. What is the intensity of your emotional state as a result of the scenario? (1---low intensity, 7---high intensity)
    \item Please explain your responses to the above slider questions.
    \item (Scenario 1 only) What would you say now was the reason you didn't succeed?
    \item (Scenario 1 only) How do you feel now about the not being able to accomplish your goal?
    \item (Scenario 1 only) Resilience: Benchmark Resilience Tool (BRT) \cite{smith2008brief} 
    \item (Scenario 2 only) What would you say now is the reason for the difficulty behind establishing your personal habit?
    \item (Scenario 2 only) How do you feel now about your ability to establish the personal habit?
    \item (Scenario 2 only) Motivation: How motivated are you to achieve this goal? (Extremely unmotivated, Unmotivated, Slightly unmotivated, Indifferent, Slightly motivated, Motivated, Extremely motivated)
    \item (Scenario 2 only) Confidence: How confident are you that you will be able to achieve this goal? (Extremely unsure, Unsure, Slightly unsure, Indifferent, Slightly confident, Confident, Extremely confident)
    \item (Scenario 2 only) Goal Commitment Measure scale \cite{hollenbeck1989investigation}
\end{itemize}

\newpage
\section{Additional Study Results}

\subsection{Demographics by Condition}
Table \ref{tab:demographics} shows the demographics breakdown for each condition.

\subsection{Clustering Results}
\label{sec:clustering}

We performed a hierarchical clustering based on participants' pre- and post-intervention responses to each scenario. Based on the clusters, we compare the control variables and demographic information of the sample population in each cluster to derive trends and insights. Below are the clusters for each scenario.

\begin{figure}[H]
    \centering
    \includegraphics[width=1\linewidth]{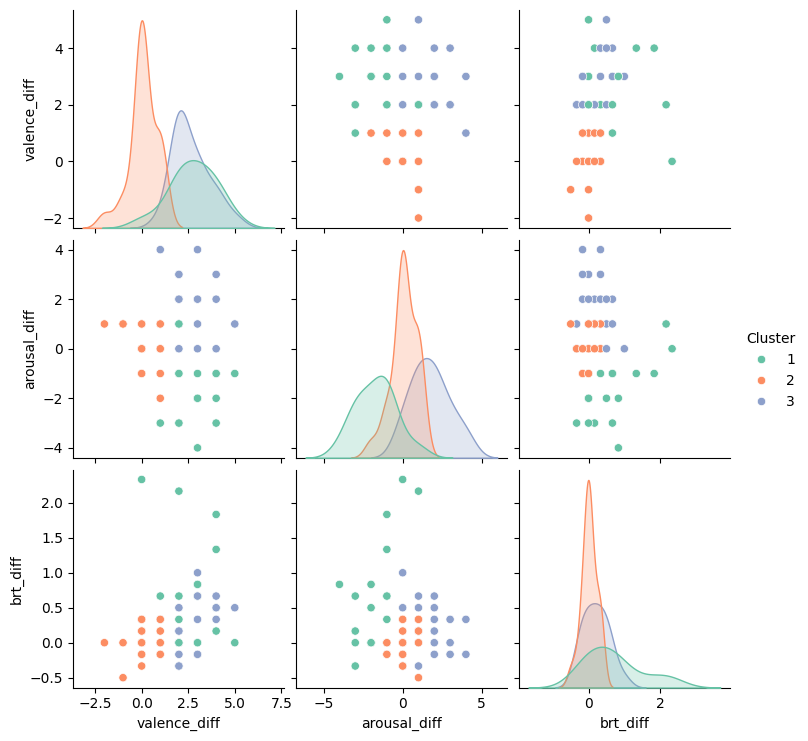}
    \caption{Pair plot visualizing the relationship between the outcomes across 3 clusters of scenario 1.}
    \label{fig:enter-label}
\end{figure}

\begin{figure}[H]
    \centering
    \includegraphics[width=1\linewidth]{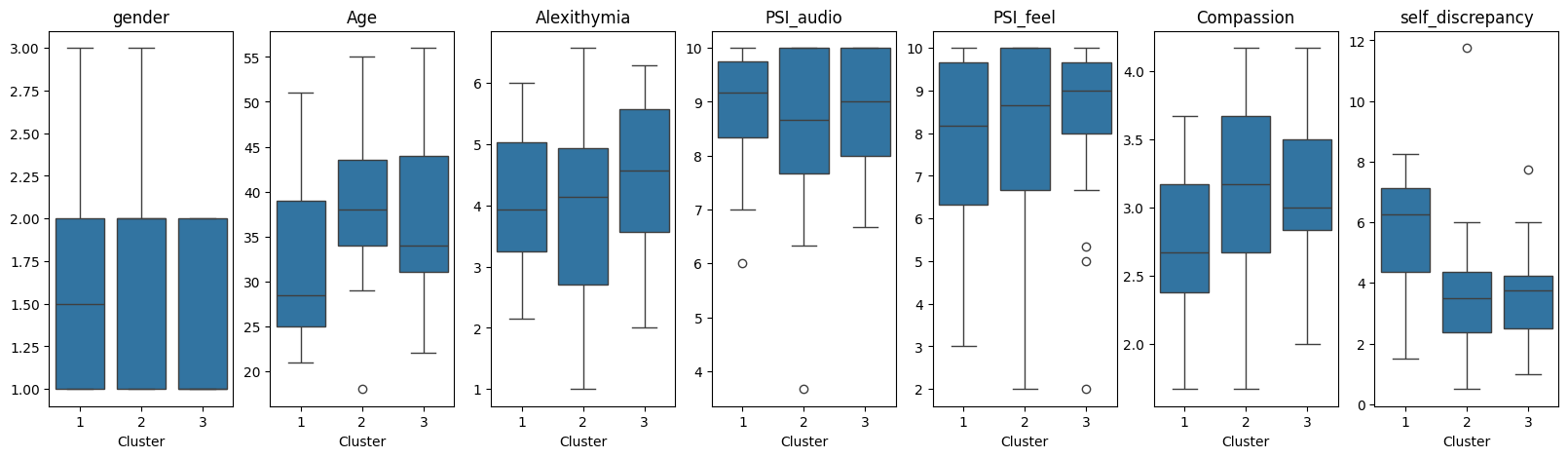}
    \caption{Demographic information of the clusters of Scenario 1.}
    \label{fig:enter-label}
\end{figure}

\begin{figure}[H]
    \centering
    \includegraphics[width=1\linewidth]{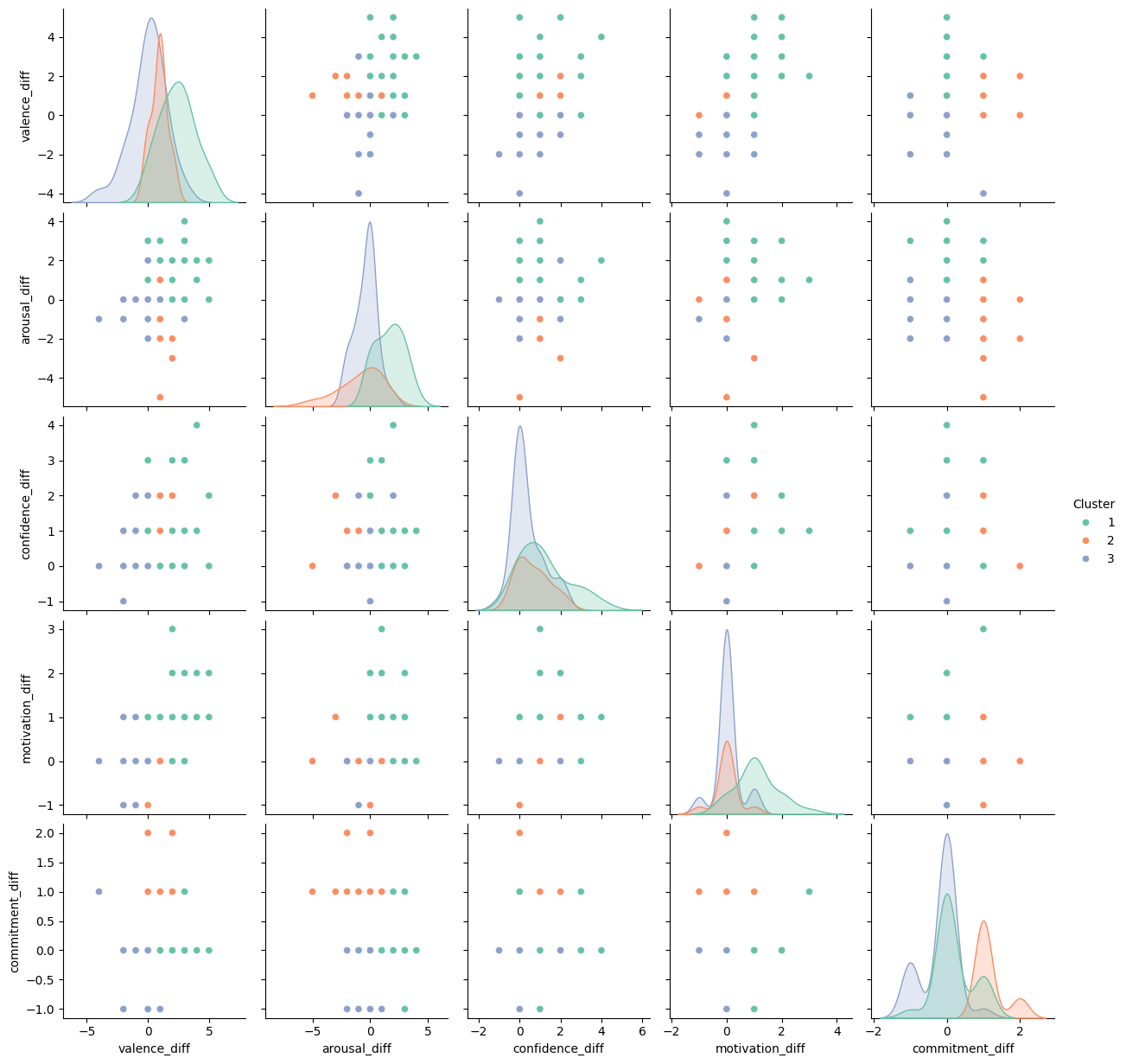}
    \caption{Pair plot visualizing the relationship between the outcomes across 3 clusters of scenario 2.}
    \label{fig:enter-label}
\end{figure}
\begin{figure}[H]
    \centering
    \includegraphics[width=1\linewidth]{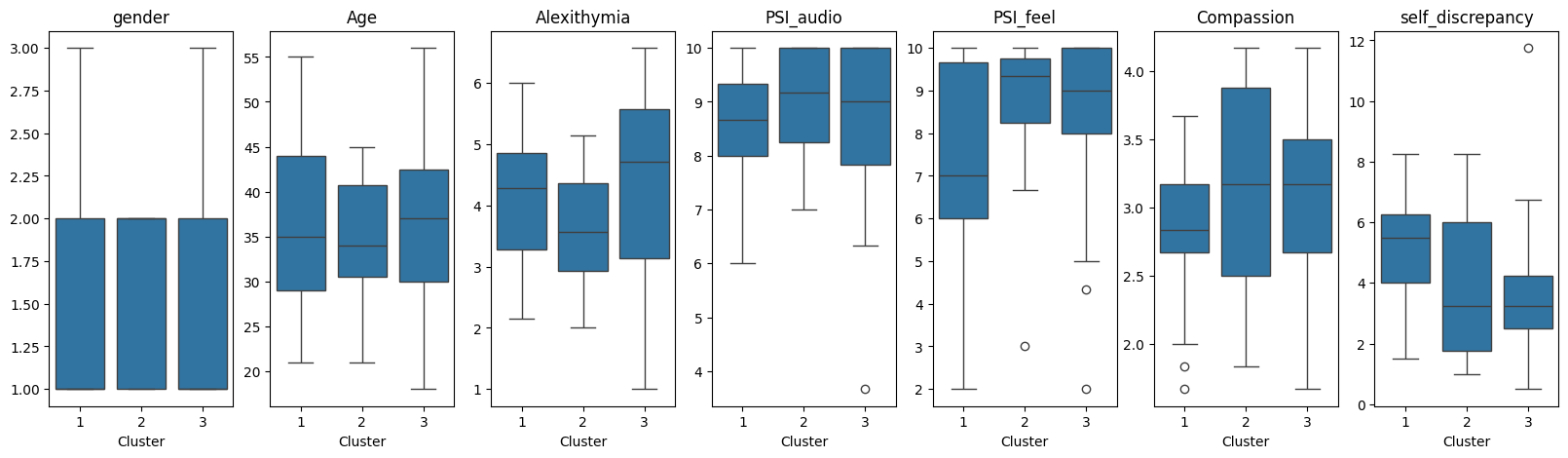}
    \caption{Demographic information of the clusters of Scenario 2.}
    \label{fig:enter-label}
\end{figure}


\end{document}